\documentclass[]{taira}

\usepackage{graphicx}
\usepackage{newtxtext}
\usepackage{newtxmath}
\usepackage{natbib}

\newcommand{\RomanNumeralCaps}[1]
\linenumbers

\usepackage{graphicx,amssymb,amsmath,color,epsfig,soul,rotating,overpic,varwidth,xcolor}

\usepackage[export]{adjustbox}
\usepackage{subfigure}
\usepackage{subfigmat}
\usepackage{algorithm}
\usepackage[noend]{algpseudocode}
\usepackage{bm}
\usepackage{xfrac}
\usepackage{multicol}
\usepackage{multirow}
\usepackage{pbox}
\usepackage{tabularx}
\usepackage{caption}

\usepackage{booktabs}
\usepackage{tikz}
\usetikzlibrary{tikzmark}

\usepackage{lipsum}

\graphicspath{{/home/jean/Documents/CODES/3Dairfoil/taperedWing/figs/}}

\definecolor{blue2}{rgb}{0, 0.4470, 0.7410}
\definecolor{red2}{rgb}{0.8500, 0.1250, 0.0480} 
\definecolor{orange2}{rgb}{0.8500, 0.3250, 0.0980} 
\definecolor{yellow2}{rgb}{0.9290, 0.6940, 0.1250}
\definecolor{purple2}{rgb}{0.4940, 0.1840, 0.5560}
\definecolor{green2}{rgb}{0.4660, 0.6740, 0.1880}
\definecolor{ltblue2}{rgb}{0.3010, 0.7450, 0.9330}
\definecolor{dkred2}{rgb}{0.6350, 0.0780, 0.1840}
\definecolor{gray2}{rgb}{0.22, 0.22, 0.3}
\definecolor{ltgray2}{rgb}{0.647, 0.647, 0.647}
\definecolor{blueIV}{rgb}{0, 0, 0.7410}
\definecolor{blueIII}{rgb}{0.2, 0.2, 0.7410}
\definecolor{blueII}{rgb}{0.4, 0.4, 0.7410}
\definecolor{blueI}{rgb}{0.7410, 0.7410, 0.7410}
\definecolor{jetVI}{rgb}{0.9763    0.9831    0.0538}
\definecolor{jetV}{rgb}{0.9264    0.7256    0.2996}
\definecolor{jetIV}{rgb}{0.4783    0.7489    0.4877}
\definecolor{jetIII}{rgb}{0.0282    0.6663    0.7574}
\definecolor{jetII}{rgb}{0.0582    0.4677    0.8589}
\definecolor{jetI}{rgb}{0.2081    0.1663    0.5292}
\definecolor{mode1}{rgb}{0 0.4470 0.7410}
\definecolor{mode2}{rgb}{0.8500 0.3250 0.0980}
\definecolor{mode3}{rgb}{0.9290 0.6940 0.1250}
\definecolor{dkgold}{rgb}{0.5930 0.5150 0.3260}
\definecolor{mode1}{rgb}{0 0.4470 0.7410}
\definecolor{mode2}{rgb}{0.8500 0.3250 0.0980}
\definecolor{mode3}{rgb}{0.9290 0.6940 0.1250}
\definecolor{dkgold}{rgb}{0.5930 0.5150 0.3260}
\definecolor{matblue}{rgb}{0.000 0.447 0.741}
\definecolor{matred}{rgb}{0.850 0.325 0.098}
\definecolor{matyellow}{rgb}{0.9290 0.6940 0.125}
\definecolor{matpurple}{rgb}{0.494 0.184 0.556}
\definecolor{matgreen}{rgb}{0.466 0.674 0.188}
\definecolor{matcyan}{rgb}{0.3010 0.7450 0.9330}
\newcommand{\bs}{\boldsymbol}

\usepackage{hyperref}
\hypersetup{
	colorlinks = true,
	urlcolor   = black,
	citecolor  = black,
}

\begin{document}
	
	\captionsetup{font=scriptsize,labelfont=scriptsize}
	
	\shorttitle{Laminar post-stall wakes of tapered swept wings} 
	\shortauthor{J. H. M. Ribeiro et al.} 
	
	\title{Laminar post-stall wakes of tapered swept wings} 
	
	\author
	{
		Jean H\'{e}lder Marques Ribeiro\aff{1}
		\corresp{\email{jeanmarques@g.ucla.edu}},
		Jacob Neal\aff{2},
		Anton Burtsev\aff{3}
		Michael Amitay\aff{2},
		Vassilios Theofilis\aff{3},
		\and
		Kunihiko Taira\aff{1}
	}
	
	\affiliation
	{
		\aff{1}
		Department of Mechanical and Aerospace Engineering, University of California, Los Angeles, CA 90095, USA
		
		\aff{2}
		Department of Mechanical, Aeronautical, and Nuclear Engineering, Rensselaer Polytechnic Institute, Troy, NY 12180, USA
		
		\aff{3}
		Department of Mechanical and Aerospace Engineering, University of Liverpool, Brownlow Hill, Liverpool L69 3GH, UK
		
		
	}
	
	\maketitle
	
	\begin{abstract}
		While tapered swept wings are widely used, the influence of taper on their post-stall wake characteristics remains largely unexplored. To address this issue, we conduct an extensive study using direct numerical simulations to characterize the wing taper and sweep effects on laminar separated wakes. We analyze flows behind NACA 0015 cross-sectional profile wings at post-stall angles of attack $\alpha=14^\circ$--$22^\circ$ with taper ratios $\lambda=0.27$--$1$, leading edge sweep angles $0^\circ$--$50^\circ$, and semi aspect ratios $sAR =1$ and $2$ at a mean-chord-based Reynolds number of $600$. 
		Tapered wings have smaller tip chord length, which generates a weaker tip vortex, and attenuates inboard downwash. This results in the development of unsteadiness over a large portion of the wingspan at high angles of attack. For tapered wings with backward-swept leading edges unsteadiness emerges near the wing tip. On the other hand, wings with forward-swept trailing edges are shown to concentrate wake shedding structures near the wing root. For highly swept untapered wings, the wake is steady, while unsteady shedding vortices appear near the tip for tapered wings with high leading edge sweep angles. For such wings, larger wake oscillations emerge near the root as the taper ratio decreases. While the combination of taper and sweep increases flow unsteadiness, we find that tapered swept wings have more enhanced aerodynamic performance than untapered and unswept wings, exhibiting higher time-averaged lift and lift-to-drag ratio. The current findings shed light on the fundamental aspects of flow separation over tapered wings in the absence of turbulent~flow~effects.

	\end{abstract}

	\section{Introduction}
	\label{sec:intro}
	
	Flow separation over aerodynamic lifting bodies has been a subject of research interest for decades, especially for small-scale air vehicles \citep{Mueller:01, Anderson:10}. To further understand the post-stall wake dynamics, it is important to analyze the influence of wing planform geometry. This characterization is challenging for high-Reynolds-number flows, due to the multiscale nature of the wakes. Nevertheless, for massively separated flows, the large vortex structures observed in higher-Reynolds-number flows are topologically analogous to the core structures in low-Reynolds-number flows \citep{HuntEtAl:JFM1978,DallmannFDR1988,Delery:AR01}. To examine the fundamental aspects of unsteady $3$-D flow separation, we study post-stall flows in the absence of turbulence \citep{Taira:JFM09,Zhang:JFM20b,Zhang:JFM20}. This characterization has been largely unexplored for low-Reynolds-number flows over tapered~wings.
	
	In aircraft design, tapered wings are used to approximate the elliptic aerodynamic loading over the wingspan. Tapered wings are more feasible to manufacture due to their less complex geometry compared to elliptic wings \citep{Prandtl:20,Mccormick:95}. The usage of tapered wings in aeronautics led to initial studies that explored the wing taper effect, especially for high-Reynolds-number flows \citep{Millikan:JAS36,Anderson:36determination,Irving:37,Soule:40design,Falkner:50}. For the laminar flow regime, the effect of wing taper on the wake dynamics is critical as the local Reynolds number is drastically reduced near the tip. For flows over wings at a chord-based Reynolds number $Re_c = \mathcal{O}(10^4)$, taper affects the aerodynamic loading with an increase in the pressure drag \citep{Traub:JA13,Traub:AIAAJ15}. For  $Re_c = \mathcal{O}(10^3)$, the aerodynamic characteristics are affected significantly by the viscous effects and the influence of wing taper on the wakes remains elusive, especially for massively separated flows.
	
	
	Post-stall wake dynamics has attracted attention of aeronautical researchers for many decades. The early efforts to understand post-stall flows over wings were performed over two-dimensional ($2$-D) spanwise homogeneous wings \citep{Abbott:59,Gaster:67,Tobak:AR82}. Valuable insights were obtained from $2$-D analysis characterizing the behavior of the separated laminar boundary layer \citep{Horton:68} and describing the relation between vortex shedding structures, adverse pressure gradient, and shear layer characteristics \citep{Pauley:JFM90}. Moreover, the emergence of wake patterns associated with 3-D separation bubbles,   as predicted in topological studies \citep{HornungPerry1984,PerryHornung1984}, was shown by global linear stability analysis to arise from self-excitation of the laminar separation bubble \citep{TheofilisHeinDallmann}.
	
	The analysis of $2$-D flows around canonical  wings continues providing fundamental insights on the effect of angle of attack and Reynolds number on the wake shedding structures \citep{Lin:AIAAJ96low,Huang:JFM01,Yarusevych:JFM09,Rossi:JFM18,Durante:CNSNS20bifurcations}. For separated flows, the increase in Reynolds number and the angle of attack yields a $3$-D flow field even around infinite and spanwise homogeneous wings \citep{BippesTurk1980,Winkelman:AIAAJ80,Schewe,Braza:JFM01,Hoarau:JFM03,PandiMittal:POF19}. In such cases, spanwise fluctuations emerge, producing $3$-D vortices in the wake, as a result of the growth of $3$-D structures associated with secondary linear instability \citep{He:JFM17}.
	
	For finite wings, $3$-D wakes result from tip effects, as a strong streamwise vortex is formed rolling up around the wing tip. \citep{Winkelman:AIAAJ80,Freymuth:AIAAJ87,Toppings:JFM22}. While turbulence has an important influence on the $3$-D wake \citep{PandiMittal:POF19}, some of the core global flow structures remain coherent over a broad range of Reynolds numbers, including the quasi-spanwise midspan shedding and the tip vortex \citep{Neal:PRF23,Pandi:JFM23}. Tip vortices induce downwash inboard over the wing, which reduces the effective angle of attack near the tip, even suppressing stall formation \citep{Dong:EF20,Toppings:JFM21} and the wake shedding for low-aspect-ratio wings \citep{Taira:JFM09,Zhang:JFM20}.  The tip vortex has been extensively studied to reveal its influence on the wake dynamics, aerodynamic forces, and pitch moments \citep{Francis:JA79,Green:JFM91,Devenport:JFM96,Pelletier:JA00,Birch:JA04,Torres:AIAAJ04,Buchholz:JFM06,Yilmaz:JFM12,Ananda:AST15,He:TCFD17}. Beyond understanding the tip vortex formation and evolution, a characterization of its instabilities has enabled the  development of control techniques that improve the aerodynamic performance around finite wings \citep{Gursul:AIAAJ18,Edstrand:JFM18b,Navrose:JFM19}.
	
	Wing sweep also has a strong influence on post-stall wake dynamics. For laminar flow regimes, a number of experimental and numerical efforts were carried out to examine the effects of backward and forward wing sweep \citep{Yen:AIAAJ07,Yen:JFE09,Zhang:JFM20b}  and identify global modes \citep{Burtsev:JFM22,Ribeiro:JFM23triglobal} that give rise to fundamental global coherent structures of flow separation around swept wings. It is noteworthy that some of the wing sweep effects on laminar post-stall flows are topologically analogous over a wide range of Reynolds numbers. For instance, for low-Reynolds-number flows over swept wings, some of the core coherent structures emerging in the near-wake, such as the ``ram's horn'' vortex and the canard leading-edge vortices, were also identified in experiments performed at higher Reynolds numbers \citep{Black:56,Breitsamter:JA01,Neal:arxiv23}. Moreover, the stabilizing effect of the sweep-induced spanwise flow on the wake structures, which significantly impacts stall characteristics, as observed at low-Reynolds-number flows \citep{Zhang:JFM20b,Ribeiro:JFM22,Ribeiro:JFM23triglobal}, is further noticed in both experiments and high-fidelity large-eddy simulations performed at a higher Reynolds number regime \citep{Harper:64,Visbal:AIAAJ19}.
	
	The aforementioned studies highlight the importance of the low-Reynolds-number post-stall wake characterization to reveal fundamental aspects of the  unsteady 3-D flow separation physics. In fact, the insights obtained from studies of post-stall laminar flows have been important to expand our knowledge of the stalled flow physics over a wide range of Reynolds numbers. Thus far, however, most studies have not considered wing taper effects on low-Reynolds number flows at high angles of attack. Only recently, a combined experimental, numerical, and theoretical effort has been initiated towards the understanding of the laminar flow over tapered wings in post-stall flow conditions \citep{Ribeiro:AIAA23,Neal:AIAA23,Burtsev:AIAA23}. Effects of taper have been analyzed for planforms with tubercles to analyze swimming of whales  \citep{Wei:AIAAJ18}, for flows over tapered cylinders \citep{Piccirillo:JFM93,Techet:JFM98,Valles:JFS02}, and for separated wakes over tapered plates \citep{Narasimhamurthy:JFM08}. For wing planforms with continuously variable chord-length over the wingspan, the delta wings have also received substantial attention \citep{Rockwell:AIAA93,Gursul:PAS05,Taira:JFM09}. For laminar post-stall flows, wing taper was studied using trapezoidal plates \citep{Huang:JFM15}. Nonetheless, there still is a lack of fundamental studies to understand the role of taper ratio, and how it interplays with leading edge (LE) and trailing edge (TE) sweep angle effects for massively separated laminar flows.
	
	For laminar separated flows, the combined effect of wing taper and sweep remains elusive. In the present work, we aim to reveal the effects of taper in the laminar wake dynamics and the influence of LE and TE sweep angles on the vortical interactions through a comprehensive campaign of direct numerical simulations of $3$-D flows over finite NACA 0015 wings. We characterize the stalled wakes of wings with backward-swept LE and forward-swept TE, identifying the combined effects of taper and sweep on the post-stall wake dynamics. Our work is organized as follows. In section \ref{sec:problem}, we present our wing planform geometry definitions and the setup for direct numerical simulations. In section \ref{sec:results}, we offer a detailed analysis and classification of the wake structures, highlighting the effects of taper and sweep on the wakes and aerodynamic forces. Finally, we conclude our study by summarizing our findings in section \ref{sec:conclusions}.
	
	\section{Problem setup}
	\label{sec:problem}
\begin{figure}
		\centering
		\begin{tikzpicture}
		\node[anchor=south west,inner sep=0] (image) at (0,0) {\includegraphics[trim=0mm 0mm 0mm 0mm, clip,width=1\textwidth]{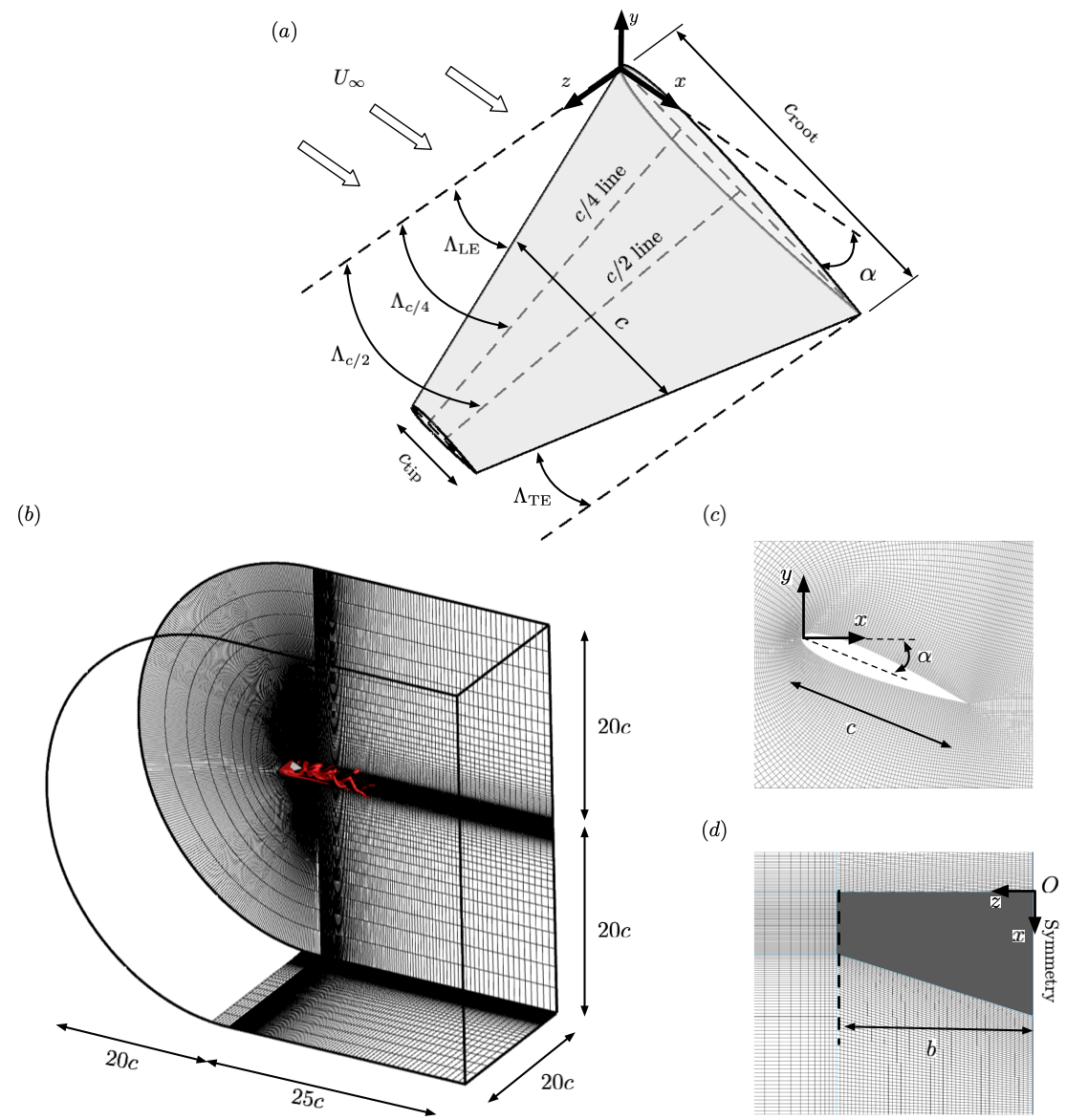}};
		\end{tikzpicture} \vspace{-6mm}
		\caption{Problem setup for tapered wings. $(a)$ Geometrical parameters shown in a wing planform with $sAR = b/c = 2$, $\alpha = 18^\circ$, $\lambda = 0.27$, and $\Lambda_\text{LE} = 18.4^\circ$. $(b)$ For a $(\lambda,\Lambda_\text{LE}) = (0.5,0^\circ)$ wing, we show the computational domain and $(c,d)$ grids with $2$-D planes at $z/c = 1$ and $y/c = -0.5$, respectively.} 
		\label{fig:setup_mesh}
	\end{figure}

	We consider laminar flows over tapered wings with a NACA 0015 cross-sectional profile. The spatial coordinates of streamwise, transverse, and spanwise directions are denoted by $(x,y,z)$, respectively. The origin is placed at the LE of the wing root, as shown in figure \ref{fig:setup_mesh}. The NACA 0015 profile is defined on the $(x,y)$ plane, which is extruded from the wing root in the spanwise direction to form the $3$-D wing. Wing taper is defined by the taper ratio $\lambda = c_\text{tip} / c_\text{root}$, where $c_\text{tip}$ and $c_\text{root}$ are tip and root chord-lengths, respectively, as shown in figure \ref{fig:setup_mesh}($a$). For all wings considered herein, the chord length decreases linearly from root to tip. The non-dimensional mean chord length $c$ at the spanwise location of $z = b/2$ is taken to be the characteristic length used to non-dimensionalize all spatial variables. The mean chord $c$ is fixed and independent of $\lambda$ for all wing planforms studied herein.
	
	The semi aspect ratio of the wings is set as $sAR = b/c = 1$ and $2$, where $b$ is the half-span length, as shown in figure \ref{fig:setup_mesh}($d$). We consider half-span wing models with symmetry imposed at the root. The angles of attack, $\alpha = 14^\circ$, $18^\circ$, and $22^\circ$, are defined  between the airfoil chord line and the streamwise direction. The present wing geometries have sharp trailing edge and straight-cut wing tip. The mean-chord-based Reynolds number is set to $Re_{c} = 600$ and the freestream Mach number is set to $M_\infty = 0.1$. Taper changes the local Reynolds number $Re_{L_c}$, defined as a function of the spanwise location \citep{Traub:AIAAJ15}. For the present study, the difference between $c_\text{tip}$ and $c_\text{root}$  accounts for a maximum variation of $60\%$ on $Re_{L_c}$ along the span, from $\min(Re_{L_c}) = 250$ to $\max(Re_{L_c}) = 950$ at the lowest taper ratio. 
	
	For tapered swept wings, the $3$-D computational setup is sheared in the chordwise direction and the LE sweep angle is defined between the $z$-direction and the LE. Tapered wings have different LE and TE sweep angles ($\Lambda_\text{LE}$ and $\Lambda_\text{TE}$, respectively), as shown in figure \ref{fig:setup_mesh}($a$). Note that the wing planform can be specified with two parameters out of the three parameters of taper ratio ($\lambda$), LE, and TE sweep. Given that $(c_\text{tip}+c_\text{root})/2 = c$, for a chosen $\lambda$, $\Lambda_\text{LE}$, and $sAR$ we have
		\begin{equation}
		\lambda = \frac{c_\text{tip}}{c_\text{root}}, \quad \frac{c_\text{root}}{c}= \frac{2}{1 + \lambda}, \quad \text{and} \quad \Lambda_\text{TE} = \arctan\left[ -\frac{2}{sAR}\left(\frac{1 - \lambda}{1 + \lambda}\right) + \tan{(\Lambda_\text{LE})} \right] \mbox{  .}
		\end{equation}
	In this work, we explore the combined effects of the LE and TE sweep angles on the wake dynamics for LE sweep angles $0 \le \Lambda_\text{LE} \le 50^\circ$ and taper ratios $0.27 \le \lambda \le 1$. The corresponding TE sweep angles take $-30^\circ \le \Lambda_\text{TE} \le 50^\circ$. Herein, negative sweep angles indicate a forward sweep, as shown in figure \ref{fig:setup_mesh}($a$), while a positive sweep angle represents a backward sweep.
	
	Traditionally in aeronautics, tapered swept wings have wing sweep angles observed with respect to the quarter-chord line \citep{Anderson:10,Anderson:36determination,Falkner:50} denoted by $\Lambda_{c/4}$, as shown in figure \ref{fig:setup_mesh}($a$). \cite{Anderson:99aircraft} considered the half-chord sweep angle $\Lambda_{c/2}$, such that aerodynamic load distribution becomes independent of the taper ratio. Straight tapered wings with $\Lambda_{c/4} = 0^\circ$ were studied by \cite{Traub:AIAAJ15}. On the other hand, \cite{Irving:37} considered the effect of the LE and TE sweep angles. For the present laminar post-stall wakes, due to the crucial role played  by the LE vortex in defining the wake characteristics \citep{Videler:Science04,Eldredge:ARFM19}, we focus on the distinct effects of $\Lambda_\text{LE}$ and $\Lambda_\text{TE}$ in our analysis and describe their influence on the wake dynamics. We note, however, that it is also possible to translate the findings reported herein with respect to the traditional quarter-chord and half-chord sweep angles, $\Lambda_{c/4}$ and $\Lambda_{c/2}$, respectively.
	
	\subsection{Direct numerical simulations}
	\label{sec:problem_dns}
	
	We conduct direct numerical simulations with a compressible flow solver \textit{CharLES} \citep{Khalighi:AIAA11,Bres:AIAAJ17}, which uses a second-order accurate finite-volume method in space with a third-order accurate total-variation diminishing Runge--Kutta scheme for time integration. The computational domain is discretized with a  C-type grid with mesh refinement near the wing and in the wake. With the origin at the airfoil LE on the symmetry plane $(x/c,\ y/c,z/c) = (0,0,0)$, the computational domain extends over $(x/c, y/c, z/c) \in [-20,25] \times  [-20,20] \times [0,20]$, which yields a maximum blockage ratio of $0.8\%$ for the wing with $\lambda = 0.27$, $sAR = 2$, and $\alpha = 22^\circ$. The computational setup is shown in figure \ref{fig:setup_mesh}($b$-$d$).
	
	We have prescribed a Dirichlet boundary condition of $(\rho, u_x, u_y, u_z, p) = (\rho_\infty, U_\infty, 0, 0, p_\infty)$ at the inlet and farfield boundaries, where $\rho$ is density, $p$ is pressure, $u_x$, $u_y$, and $u_z$ are velocity components in $x$, $y$, and $z$ directions respectively. The subscript $\infty$ denotes the freestream values. A symmetry boundary condition is prescribed along the root plane, $z/c = 0$. We have evaluated the applicability of the root-symmetry boundary condition by conducting DNS of flows over full wing configurations, without root-symmetry, for wings at $\alpha = 22^\circ$ and $\lambda = 0.27$ and $1$. For both wings, we note that the wake exhibits root-concentrated vortex shedding and remains symmetric with respect to the wing root over large computational times. 
	
	A no-slip adiabatic boundary condition is set on the airfoil surface. For vortical structures to convect out of the domain, a sponge layer is applied over $x/L_c \in [15,25]$ with the target state being the running time-averaged state over $5$ convective time units \citep{Freund:AIAAJ97}. Simulations start from uniform flow and are performed with a constant acoustic Courant-Friedrichs-Lewy (CFL) number of $1$ until transients are washed out of the computational domain. The time to flush out the transients varies depending on the wing planform and angle of attack, generally ranging from $50$ to $300$ convective time units. After the transients are washed out the domain, flows are simulated with a constant time step defined such that CFL is smaller than one. Flow statistics are collected for $100$ to $300$ convective time units, depending on the flow field characteristics and spectral content to ensure convergence. A detailed discussion on verification is provided in appendix \ref{sec:verification}.
	
	\section{Results}
	\label{sec:results}
	
	\subsection{Overview of tapered wing wakes}
	\label{sec:overview}
	
	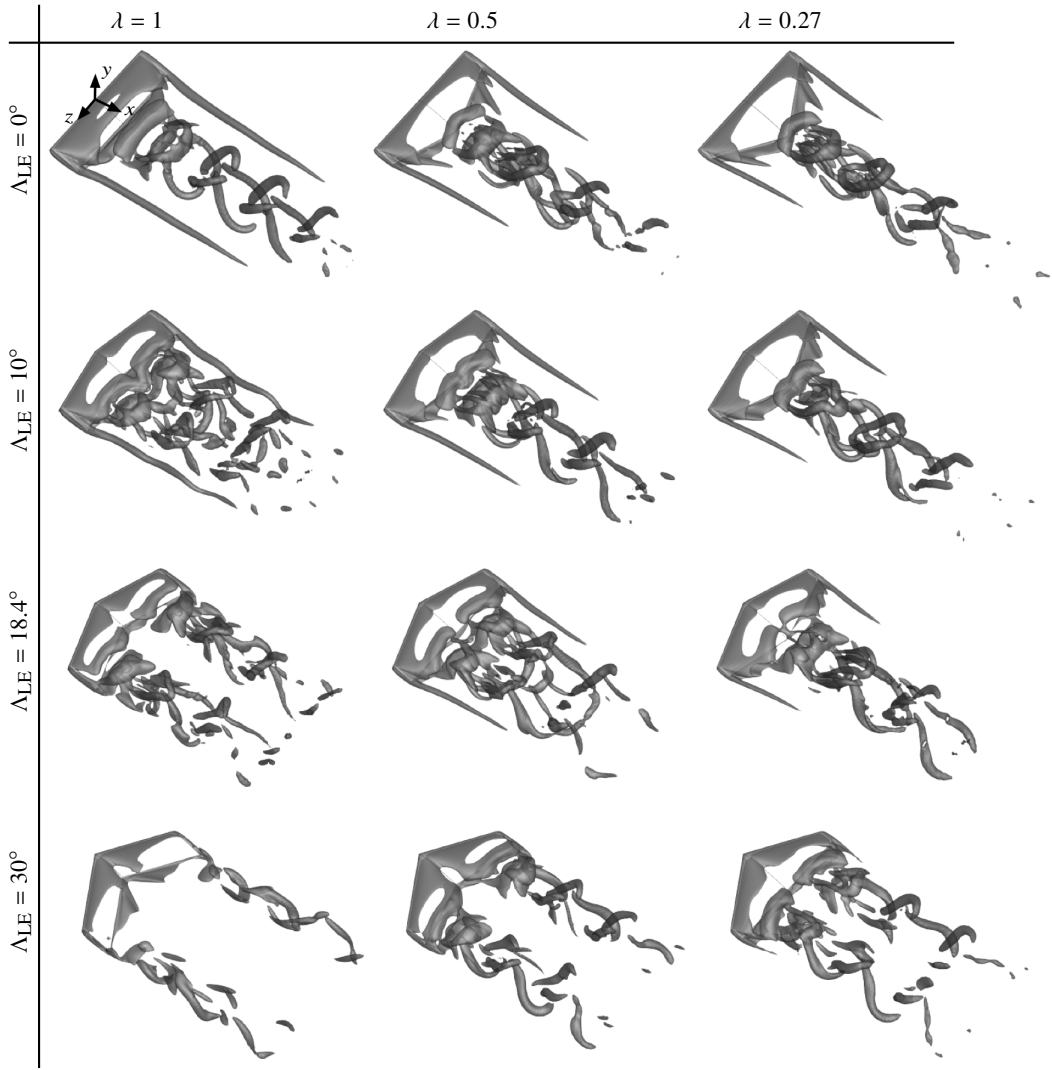
\begin{figure}
		\begin{tikzpicture}
		\node[anchor=south west,inner sep=0] (image) at (0,0) {\includegraphics[trim=0mm 0mm 0mm 0mm, clip,width=1\textwidth]{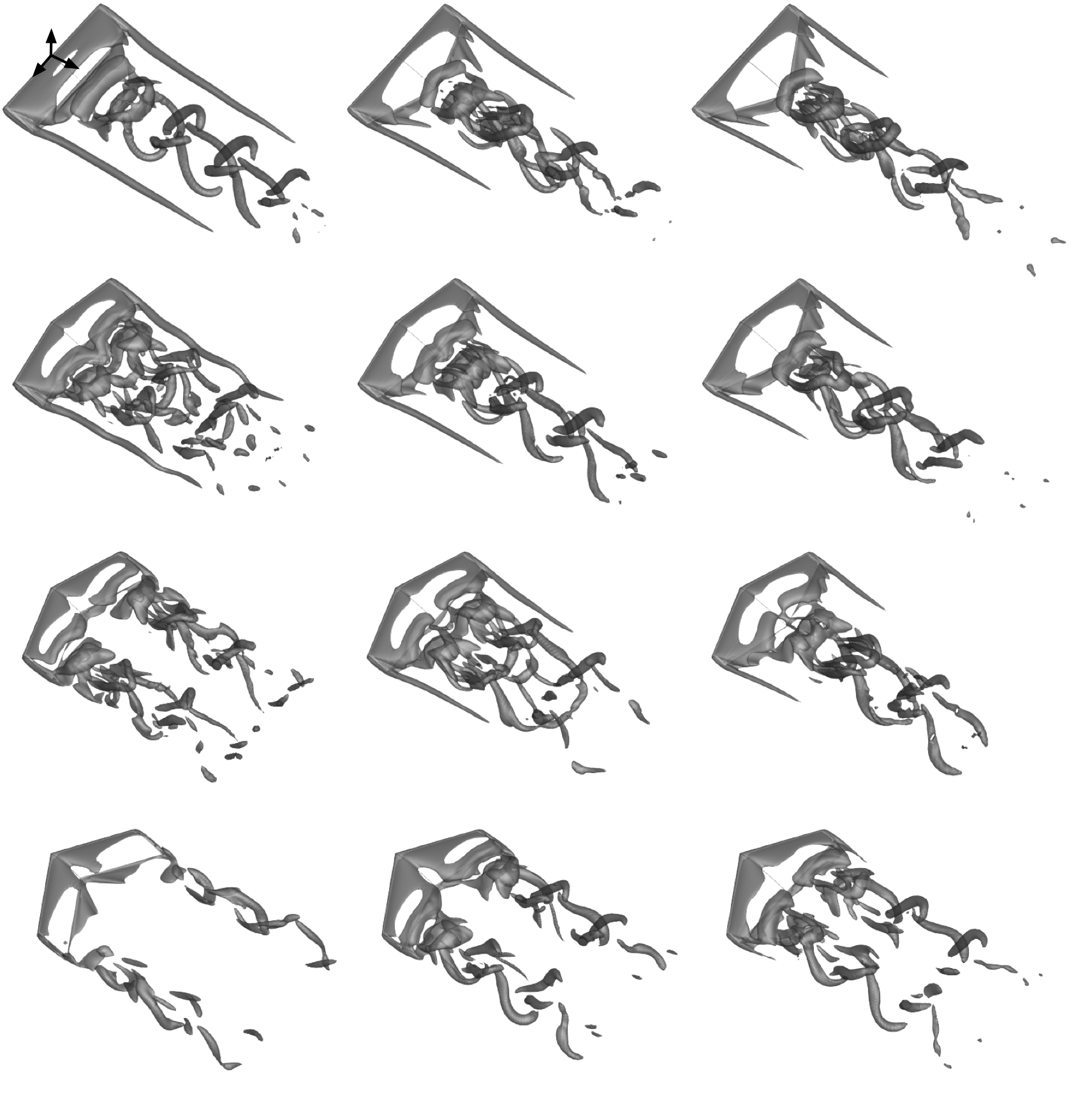}};
		\footnotesize
		\draw[black, thick] (-0.1,0.5) -- (-0.1,14.5);
		\draw[black, thick] (-0.5,14.0) -- (12.0,14.0);
		\node[align=left] at (-0.35,12.6) {\rotatebox{90}{$\Lambda_\text{LE} = 0^\circ$}};
		\node[align=left] at (-0.35,9.3) {\rotatebox{90}{$\Lambda_\text{LE} = 10^\circ$}};
		\node[align=left] at (-0.35,6.0) {\rotatebox{90}{$\Lambda_\text{LE} = 18.4^\circ$}};
		\node[align=left] at (-0.35,2.7) {\rotatebox{90}{$\Lambda_\text{LE} = 30^\circ$}};
		\node[align=left] at (1.2,14.3) {\rotatebox{0}{$\lambda = 1$}};
		\node[align=left] at (5.5,14.3) {\rotatebox{0}{$\lambda = 0.5$}};
		\node[align=left] at (9.7,14.3) {\rotatebox{0}{$\lambda = 0.27$}};
		\scriptsize
		\node[align=left] at (1.1,13.1) {\rotatebox{0}{$x$}};
		\node[align=left] at (0.8,13.6) {\rotatebox{0}{$y$}};
		\node[align=left] at (0.3,13.0) {\rotatebox{0}{$z$}};
		\end{tikzpicture} \vspace{-8mm}
		\caption{Instantaneous flows around tapered wings with $sAR = 2$, $0.27 \le \lambda \le 1$, and $0^\circ \le \Lambda_\text{LE} \le 30^\circ$, at $\alpha = 18^\circ$ visualized using gray-colored isosurfaces of $Q = 1$.} 
		\label{fig:qcritOverview}
	\end{figure}
	In figure \ref{fig:qcritOverview}, we present instantaneous post-stall flows over tapered wings, which exhibit a rich diversity of wake structures through the combined effects of LE and TE sweep. Taper effects on laminar separated flows are entwined with the effects of LE and TE sweep angles. However, by studying straight tapered wings, that is, wings with $\Lambda_{c/2}$ and $\Lambda_{c/4}$ approximately zero,  we can distinguish the effects of taper from other geometrical~parameters.
	
	For instance, let us  explore the flows over wings with $(\lambda,\Lambda_\text{LE}) = (1,0^\circ)$ and compare them to the wake structures around $(\lambda,\Lambda_\text{LE}) =(0.27, 10^\circ)$ wings; these flows have  $\Lambda_{c/4} = 0^\circ$ and $1.8^\circ$, respectively. For the tapered wing, we note a significant reduction of the tip vortex length caused by the smaller $c_\text{tip}$. The downstream root shedding, however, exhibits similar hairpin-like structures for both wings. In the near-wake region, for the tapered wing, the spatial spanwise flow fluctuations emerge near the root, over the spanwise vortex on the suction side. Such wake oscillations are absent in the vortical structure that forms over the untapered wing. 
	
	We can further explore the distinct taper effects on the wake dynamics by considering wings with $\Lambda_{c/2} \approx 0^\circ$, as shown for the similar flow patterns that develop at the root region for $(\lambda,\Lambda_\text{LE}) = (1,0^\circ)$ and $(0.27, 18.4^\circ)$ wings. Here, with a lower taper ratio, tip vortices are considerably weakened when compared to the structures near the free end of the untapered wing. The vortical roll structures emerging over the tapered wing appear slanted and aligned with $\Lambda_\text{LE}$, suggesting that the LE sweep angle plays an important role to define the behavior of the near-wake shedding structures.
	
	For tapered wings, the backward-swept LE effect can be observed by fixing the $\Lambda_\text{TE} = 0^\circ$ while the LE is swept backwards with $\Lambda_\text{LE} = 18.4^\circ$ and $30^\circ$ for $\lambda = 0.5$ and $0.27$, respectively. For such wings, taper shifts the wake shedding structures closer to the wing tip. An opposite effect is shown in the top row of figure \ref{fig:qcritOverview}, for flows over forward-swept TE wings. These planforms have fixed $\Lambda_\text{LE} = 0^\circ$, while $\Lambda_\text{TE} = -18.4^\circ$ and $-30^\circ$ for $\lambda = 0.5$ and $0.27$, respectively. For these cases, we observe that taper reduces the tip vortex length and changes the topology of the root shedding structures. Let us further study the taper effect for highly swept wings, shown at the bottom row of figure \ref{fig:qcritOverview}, with a fixed $\Lambda_\text{LE} = 30^\circ$, while $\Lambda_\text{TE} = 13.7^\circ$ and $0^\circ$ for $\lambda = 0.5$ and $0.27$, respectively. Here, taper increases the amplitude of wake oscillations. We further detail the discussions on the effects of taper, LE, and TE sweep in section \ref{sec:LEandTEsweep}.
	
	The variety of wake structures that appear around tapered wings, as seen in figure \ref{fig:qcritOverview}, calls for a proper characterization of the wake dynamics that associates its behavior with the wing planform geometry. The above discussions suggest that taper affects the location where unsteadiness emerges and the characteristics of the vortical structures. In the following section, we provide a map that characterized the wakes of tapered wings.
	
	\subsection{Wake classification and aerodynamic forces}
	\label{sec:wakecharacterization}
	
	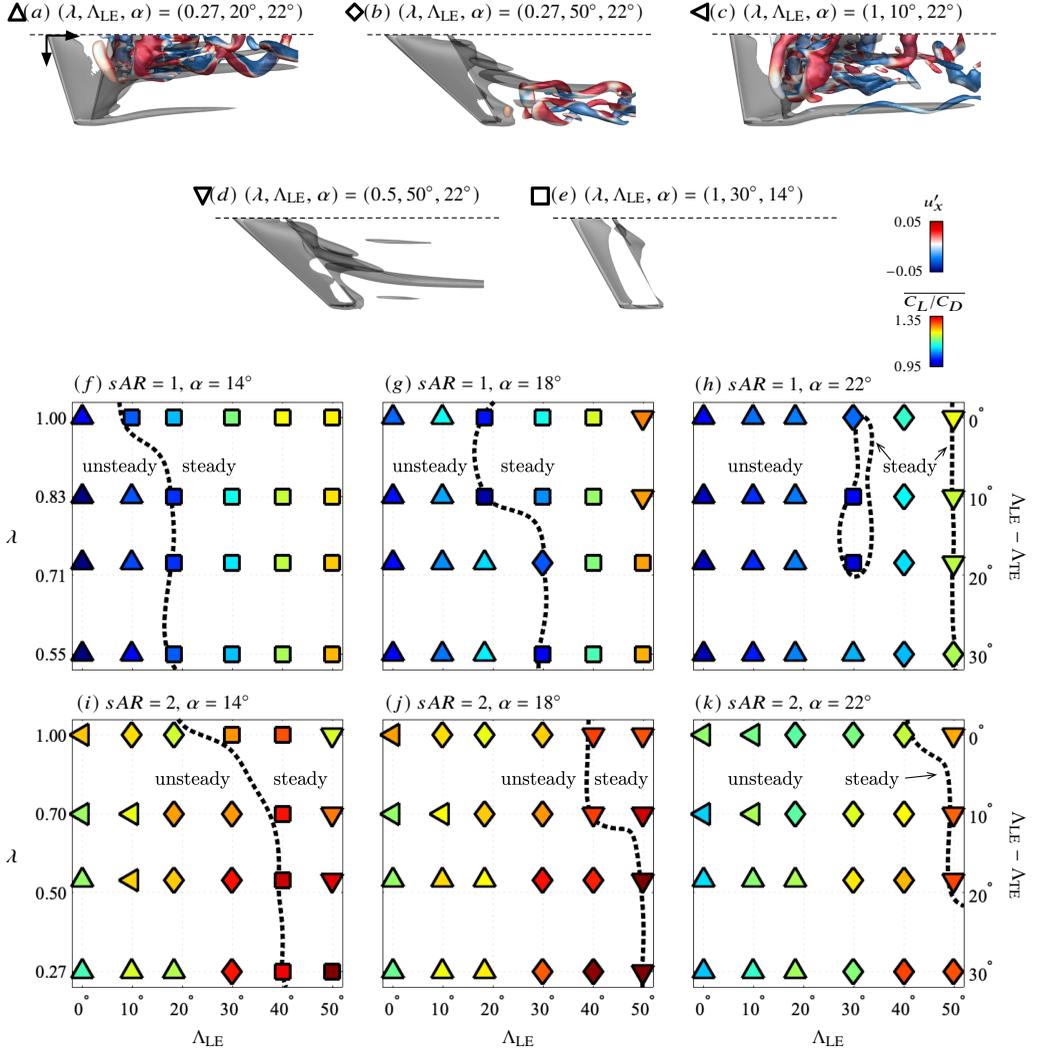
\begin{figure}
		\centering
		\begin{tikzpicture}
		\node[anchor=south west,inner sep=0] (image) at (0,0) {\includegraphics[trim=0mm 0mm 0mm 0mm, clip,width=1.0\textwidth]{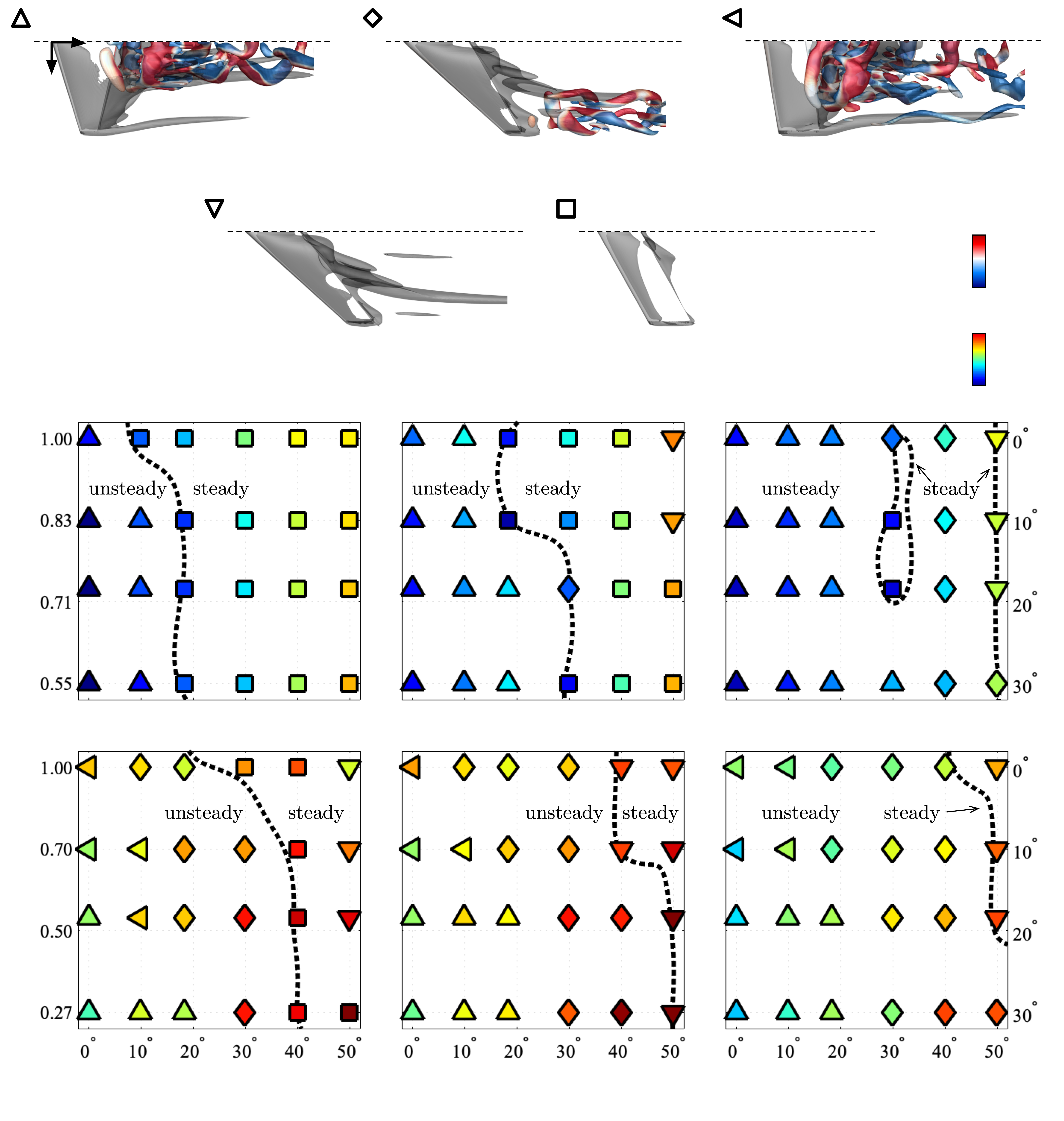}};
		\scriptsize
		\node[align=left] at (2.2,9.1) {\rotatebox{0}{$(f)$ $sAR = 1$, $\alpha = 14^\circ$}};
		\node[align=left] at (6.3,9.1) {\rotatebox{0}{$(g)$ $sAR = 1$, $\alpha = 18^\circ$}};
		\node[align=left] at (10.4,9.1) {\rotatebox{0}{$(h)$ $sAR = 1$, $\alpha = 22^\circ$}};
		\node[align=left] at (2.2,4.9) {\rotatebox{0}{$(i)$ $sAR = 2$, $\alpha = 14^\circ$}};
		\node[align=left] at (6.3,4.9) {\rotatebox{0}{$(j)$ $sAR = 2$, $\alpha = 18^\circ$}};
		\node[align=left] at (10.4,4.9) {\rotatebox{0}{$(k)$ $sAR = 2$, $\alpha = 22^\circ$}};
		\node[align=left] at (0.2,7.10) {\rotatebox{0}{$\lambda$}};
		\node[align=left] at (0.2,2.90) {\rotatebox{0}{$\lambda$}};
		\node[align=left] at (13.5,7.10) {\rotatebox{270}{$\Lambda_\text{LE}-\Lambda_\text{TE}$}};
		\node[align=left] at (13.5,2.90) {\rotatebox{270}{$\Lambda_\text{LE}-\Lambda_\text{TE}$}};
		\node[align=left] at (2.8,0.5) {\rotatebox{0}{$\Lambda_\text{LE}$}};
		\node[align=left] at (6.9,0.5) {\rotatebox{0}{$\Lambda_\text{LE}$}};
		\node[align=left] at (11.0,0.5) {\rotatebox{0}{$\Lambda_\text{LE}$}};
		\node[align=left] at (2.2,14.0) {\rotatebox{0}{$(a)$ $(\lambda,\Lambda_\text{LE},\alpha) = (0.27,20^\circ,22^\circ)$}};
		\node[align=left] at (6.7,14.0) {\rotatebox{0}{$(b)$  $(\lambda,\Lambda_\text{LE},\alpha) = (0.27,50^\circ,22^\circ)$}};
		\node[align=left] at (11.1,14.0) {\rotatebox{0}{$(c)$ $(\lambda,\Lambda_\text{LE},\alpha) = (1,10^\circ,22^\circ)$}};
		\node[align=left] at (4.6,11.6) {\rotatebox{0}{$(d)$ $(\lambda,\Lambda_\text{LE},\alpha) = (0.5,50^\circ,22^\circ)$}};
		\node[align=left] at (9.0,11.6) {\rotatebox{0}{$(e)$ $(\lambda,\Lambda_\text{LE},\alpha) = (1,30^\circ,14^\circ)$}};
		\tiny
		\node[align=left] at (12.4,11.5) {\rotatebox{0}{$u_x^\prime$}};
		\node[align=left] at (12.05,11.25) {\rotatebox{0}{$0.05$}};
		\node[align=left] at (12.0,10.60) {\rotatebox{0}{$-0.05$}};
		\node[align=left] at (12.4,10.2) {\rotatebox{0}{$\overline{C_L/C_D}$}};
		\node[align=left] at (12.05,9.95) {\rotatebox{0}{$1.35$}};
		\node[align=left] at (12.05,9.35) {\rotatebox{0}{$0.95$}};
		\end{tikzpicture} \vspace{-10mm}
		\caption{Classification of laminar flows over tapered wings into five distinct wake patterns ($a$-$e$) shown for $sAR = 2$ wings visualized with time-averaged $\overline{Q} = 1$ in gray and instantaneous $Q^\prime = 0.2$ colored by $u_x^\prime$. Classification map for ($f$-$h$) $sAR = 1$ and ($i$-$k$) $sAR = 2$ colored by $\overline{C_L/C_D}$. Black dashed lines mark transition from steady to unsteady flows.} 
		\label{fig:wakeClassification}
	\end{figure}
	
	We now classify the flow patterns with respect to the wing geometry. Our criterion is based on the examination of the flow characteristics downstream of the airfoil on a $2$-D plane at $x/c = 4$, where we identify the spatial location of maximum time-averaged $\overline{Q}$ and the maximum fluctuating component of $Q^\prime = Q - \overline{Q}$, where $Q$ is the second invariant of the velocity gradient tensor used to identify the vortical structures \citep{Hunt:CTR88,Jeong:JFM95}. Maximum $\overline{Q}$ and $Q^\prime$ located between $0 \le z/(c\ sAR) < 0.5$ are labeled root dominant, while points with maximum $\overline{Q}$ or $Q^\prime$ between $0.5 \le z/(c\ sAR) \le 1$ are named tip-dominant. We consider the flow as steady when the maximum fluctuating value of $Q^\prime$ is smaller than $0.1$ at $x/c=4$. Using the root and tip locations of $\overline{Q}$ and $Q^\prime$, we classify their wakes  into $3$ unsteady and $2$ steady regimes, as shown in figure \ref{fig:wakeClassification}, where the steady-unsteady threshold (black dotted line) is computed via biharmonic spline interpolation. We further verify our classification criterion by carefully inspecting the flow fields. Instantaneous flow fields for all tapered wings shown in figure \ref{fig:wakeClassification} are provided in the appendix \ref{sec:flowviz} using isosurfaces of $Q = 1$ colored by streamwise velocity $u_x$.

	The first flow regime (${\boldsymbol{\triangle}}$) is composed of tapered wings wakes that have both maximum $\overline{Q}$ and $Q^\prime$ found near the root region. Such wakes appear for tapered wings with low LE sweep angles. For such wings, the tip vortex tends to be short in length and the taper and forward-swept TE effects concentrate shedding at the wing root, as shown in figure \ref{fig:wakeClassification}$(a)$ for $(\lambda,\Lambda_\text{LE}) = (0.27, 20^\circ)$ at $\alpha = 22^\circ$. The second flow regime of unsteady wakes ({\Large ${\boldsymbol{\diamond}}$}) occurs when both maximum $\overline{Q}$ and $Q^\prime$ are found over the tip region. Such wakes are observed around tapered wings over a broad range of $\lambda$ values, being present for wings with high LE sweep angles. The flow over such wings often exhibits hairpin-like vortices downstream in the wake aligned with the wing tip, as shown in figure \ref{fig:wakeClassification}$(b)$ for  $(\lambda,\Lambda_\text{LE}) = (0.27, 50^\circ)$  at $\alpha = 22^\circ$. 
	
	The third flow regime of unsteady wakes (${\boldsymbol{\triangleleft}}$) around tapered wings presents maximum $\overline{Q}$ at the wing tip with maximum $Q^\prime$ at the root. This wake characteristic is often present for slightly tapered and swept wings, that is, wings with high $\lambda$ and low LE sweep angles. Such wings exhibit a distinct tip vortex formation, at the location of the maximum $\overline{Q}$, and wake shedding near the root. On some occasions, the tip vortex exhibits weak unsteady flow oscillations, as shown in figure \ref{fig:wakeClassification}$(c)$ for $(\lambda,\Lambda_\text{LE}) = (1, 10^\circ)$ at $\alpha = 22^\circ$, while the most energetic vortices are generally observed over the root region. 
	
	There are two distinct flow regimes of steady wakes shown herein, as seen in figure \ref{fig:wakeClassification}$(d,e)$. The first one (${\boldsymbol{\triangledown}}$) is comprised of wakes with a steady streamwise vortex that develops into the wake. Such flows are mainly exhibited around highly swept $sAR = 2$ wings with high and moderate taper ratios, $\lambda \ge 0.5$, as shown in figure \ref{fig:wakeClassification}$(d)$ for $(\lambda,\Lambda_\text{LE}) = (0.5, 50^\circ)$ at $\alpha = 22^\circ$. The second steady wakes regime (${\boldsymbol{\square}}$) is comprised of flows with no significant wake structures, with maximum $\overline{Q} \le 0.1$ in the wake and are commonly observed for $sAR = 1$ wings, as shown in figure \ref{fig:wakeClassification}$(e)$ for a $(\lambda,\Lambda_\text{LE}) = (1, 30^\circ)$ wing at $\alpha = 14^\circ$, and for $sAR = 2$ wings at lower angles of attack, low taper ratios, and high LE sweep angles.
	
	In figure \ref{fig:wakeClassification}($f$-$k$), we present the classification for all wings studied herein. For $sAR = 1$ wings, whose classification is shown in figure \ref{fig:wakeClassification}($f$-$h$), there are fewer changes in wake class, when compared to $sAR = 2$ wings. For the higher aspect ratio wings, with a fixed $\Lambda_\text{LE}$, we often notice $2$ or $3$ distinct classes of wake behavior as $\lambda$ changes. On the other hand, for $sAR = 1$ wings, the same class for all $\lambda$ is observed regularly. Taper effects become increasingly important for $sAR = 2$ wings to alter their wake characteristics, as shown in figure \ref{fig:wakeClassification}($i$-$k$), not only affecting the steady-unsteady wake behavior, but also producing distinct wakes as a function of $\lambda$. For $sAR = 2$ wings with high LE sweep angles, the transition from steady to unsteady wakes is dependent on $\lambda$. Generally, untapered wings with high LE sweep angle wakes remain steady, while unsteady flow structures emerge in the wakes of tapered swept wings. 
	
	\begin{figure}
		\centering
		\begin{tikzpicture}
		\node[anchor=south west,inner sep=0] (image) at (0,0) {\includegraphics[trim=0mm 0mm 0mm 0mm, clip,width=1.0\textwidth]{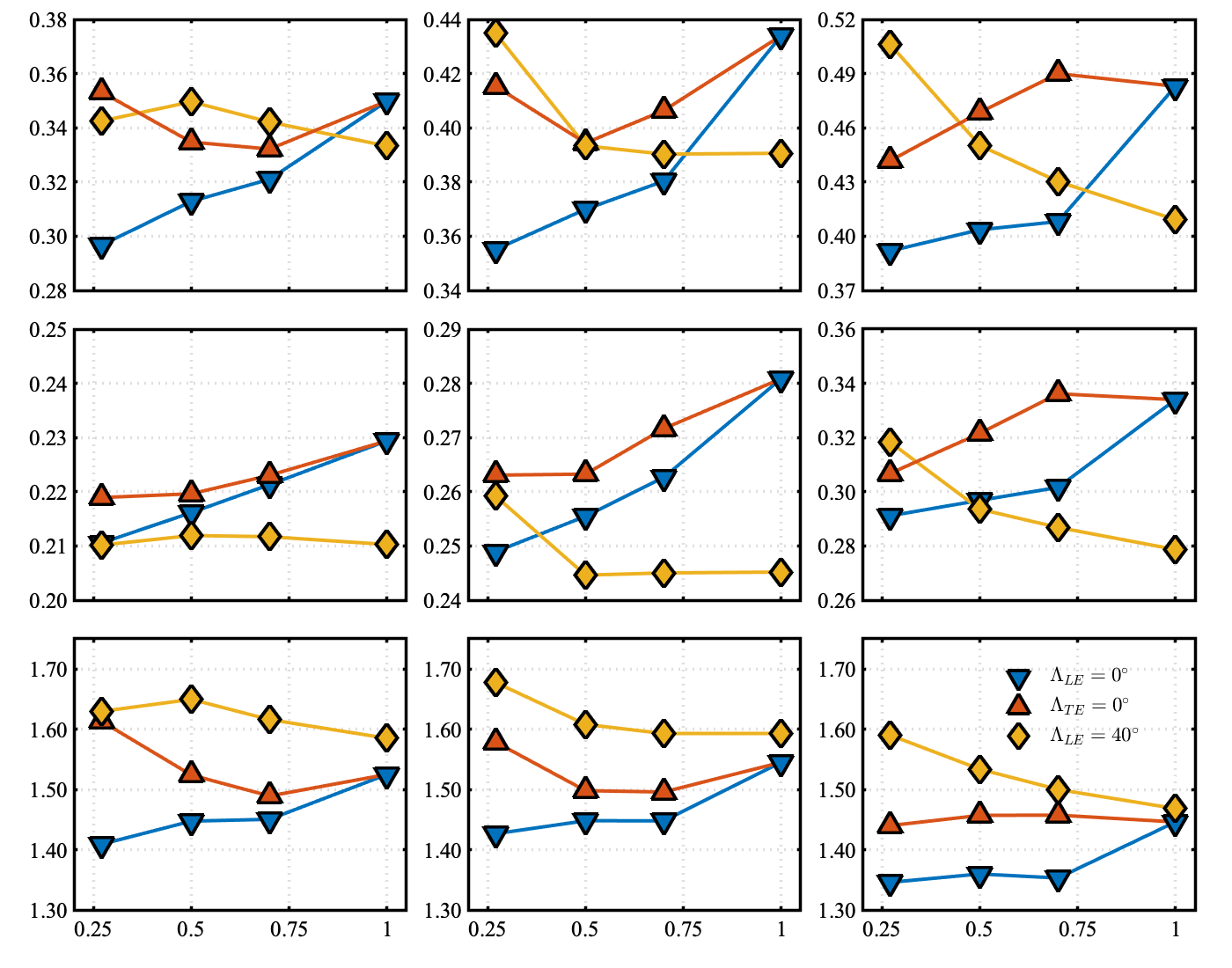}};
		\scriptsize
		\node[align=left] at (1.4,10.7) {\rotatebox{0}{$(a)$ $\alpha = 14^\circ$}};
		\node[align=left] at (5.7,10.7) {\rotatebox{0}{$(b)$ $\alpha = 18^\circ$}};
		\node[align=left] at (10,10.7) {\rotatebox{0}{$(c)$ $\alpha = 22^\circ$}};
		\node[align=left] at (0.0,9.00) {\rotatebox{90}{$\overline{C_L}$}};
		\node[align=left] at (0.0,5.50) {\rotatebox{90}{$\overline{C_D}$}};
		\node[align=left] at (0.0,1.90) {\rotatebox{90}{$\overline{C_L/C_D}$}};
		\node[align=left] at (2.6,0.0) {\rotatebox{0}{$\lambda$}};
		\node[align=left] at (7.0,0.0) {\rotatebox{0}{$\lambda$}};
		\node[align=left] at (11.3,0.0) {\rotatebox{0}{$\lambda$}};
		\end{tikzpicture} \vspace{-7mm}
		\caption{Time-averaged lift, drag, and lift-to-drag coefficients, $\overline{C_L}$, $\overline{C_D}$, and $\overline{C_D}$, respectively, for $sAR = 2$ tapered wings with $0.27 \le \lambda \le 1$ at $14^\circ \le \alpha \le 22^\circ$. Blue downward-pointing triangles), wings with unswept LE and forward-swept TE; red upward-pointing triangles, wings with backward-swept LE and unswept TE; yellow diamonds, wings with $\Lambda_\text{LE} = 40^\circ$.} 
		\label{fig:ClandCd}
	\end{figure}
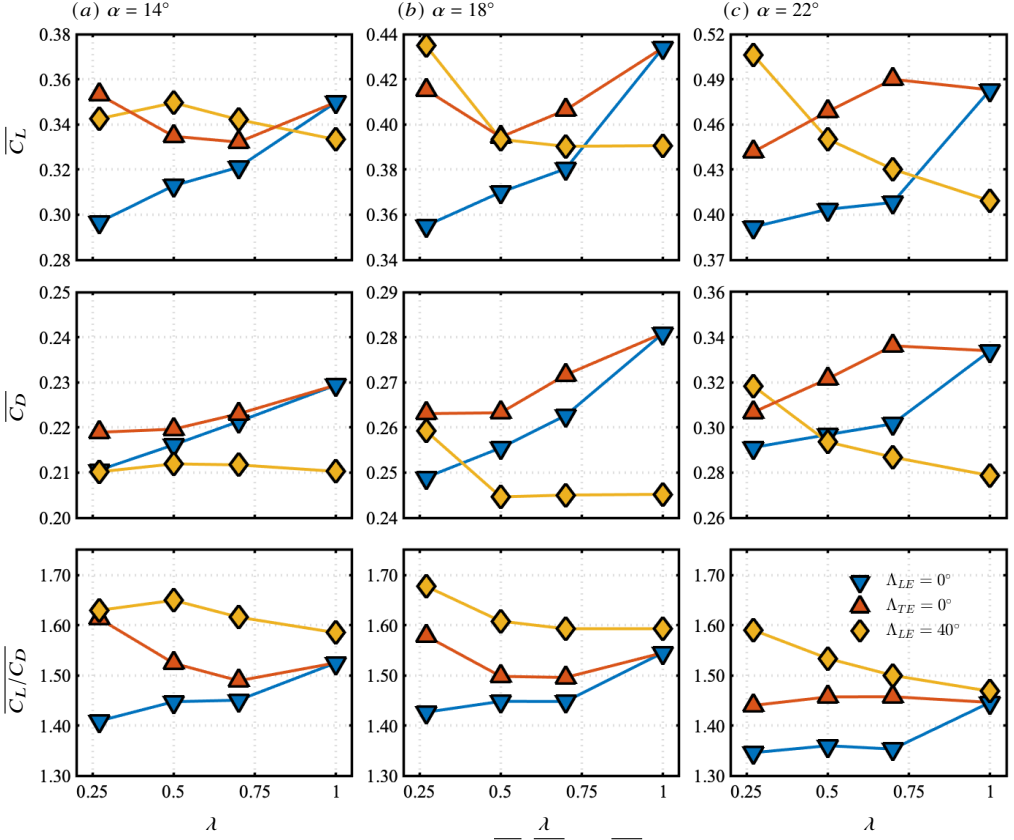
	
	The combination LE sweep angle increase and taper ratio decrease is shown to promote wake unsteadiness. In addition, tapered wings with high LE sweep angles exhibit enhanced aerodynamic performance when compared to untapered and unswept wings, as shown in figure \ref{fig:wakeClassification}($f$-$k$). To visualize this trend, each symbol associated with a wake class is colored by the time-averaged lift-to-drag ratio, $\overline{C_L/C_D}$, showing that the higher lift-to-drag ratio coefficients, for all $sAR$ and $\alpha$ combination, appear for the wings with lower $\lambda$ and higher backward-swept LE. Here, the aerodynamic forces are reported with lift and drag coefficients defined as
	\begin{equation}
	C_L = \frac{F_y}{\frac{1}{2}\rho U_\infty^2 b c} \quad \text{and} \quad \quad C_D = \frac{F_x}{\frac{1}{2}\rho U_\infty^2 b  c}  \mbox{  ,}
	\label{eq:cd_cl}
	\end{equation} 
	where $F_x$ and $F_y$ are the $x$ and $y$ components of the force on the wing, respectively. Furthermore, we study the aerodynamic loads through the time-averaged $\overline{C_L}$, $\overline{C_D}$, and $\overline{C_L/C_D}$ for selected $sAR = 2$ wings, as shown in figure \ref{fig:ClandCd}, to reveal the influence of taper and sweep on the aerodynamic forces. The blue symbols present the aerodynamic loads for tapered wings with unswept LE and forward-swept TE. The red symbols show the results for tapered wings with backward-swept LE and unswept TE, while the yellow symbols represent tapered wings with $\Lambda_\text{LE} = 40^\circ$. 
	
	The effects of wing taper on the wakes and aerodynamic forces are strongly dependent on the combination of taper and sweep angle. Let us start from the untapered and unswept wings, marked by blue downward-pointing triangles at $\lambda = 1$. The flow fields around these wings are characterized by root shedding and a strong tip vortex, as seen in figure \ref{fig:qcritOverview}. While keeping the LE unswept, the TE is forward-swept for low $\lambda$. In the wake, such taper produces a concentration of both steady and fluctuating wake structures near the root, as shown in figure \ref{fig:wakeClassification}($i$-$k$). A root-concentrated wake with a small tip vortex substantially decreases $\overline{C_L}$, $\overline{C_D}$, and $\overline{C_L/C_D}$ with $\lambda$. It is noteworthy that at the same $\lambda$, tapered wings have backward-swept LE and unswept TE exhibit a higher $\overline{C_L}$ and $\overline{C_L/C_D}$, as shown in figure \ref{fig:ClandCd}. For such wings, we recall that taper shifts wake structures toward the tip region, as shown in figure \ref{fig:qcritOverview}.
	
	Backward-swept LE enhances the aerodynamic efficiency of tapered wings in post-stall laminar flow conditions. This LE-sweep-induced improvement in aerodynamic loads also occurs for tapered wings with high LE sweep angles. For instance, untapered swept wings present lower values of $\overline{C_L}$ for all angles of attack, as shown by the diamond-shaped yellow symbols for $\lambda = 1$ in figure \ref{fig:ClandCd}, while significantly reducing wake oscillations, as shown in figures \ref{fig:qcritOverview} and \ref{fig:wakeClassification}($d$-$f$). 
	
	For $sAR = 2$ wings with high LE sweep at high incidence $\alpha \ge 18^\circ$, taper causes a change in the wake regime, as shown in figure \ref{fig:wakeClassification}($i$-$k$). For such wings, wake shedding emerges near the wing tip. The change in wake flow regime for tapered wings at high incidence causes an increase in $\overline{C_L/C_D}$. At lower incidence, $\alpha = 14^\circ$, the wake regime remains steady without noticeable vortices for tapered wings and the aerodynamic forces remain fairly constant for all $\lambda$, as shown in figure \ref{fig:ClandCd}$(a)$. Let us now describe in detail the taper effects on the wake characteristics in the following section and relate the wakes to the lift distribution over the wing.
	
	\subsection{Wake characteristics}
	\label{sec:LEandTEsweep}

	\subsubsection{Tapered wings with unswept LE and forward-swept TE}
	\label{sec:TEsweep}
	
	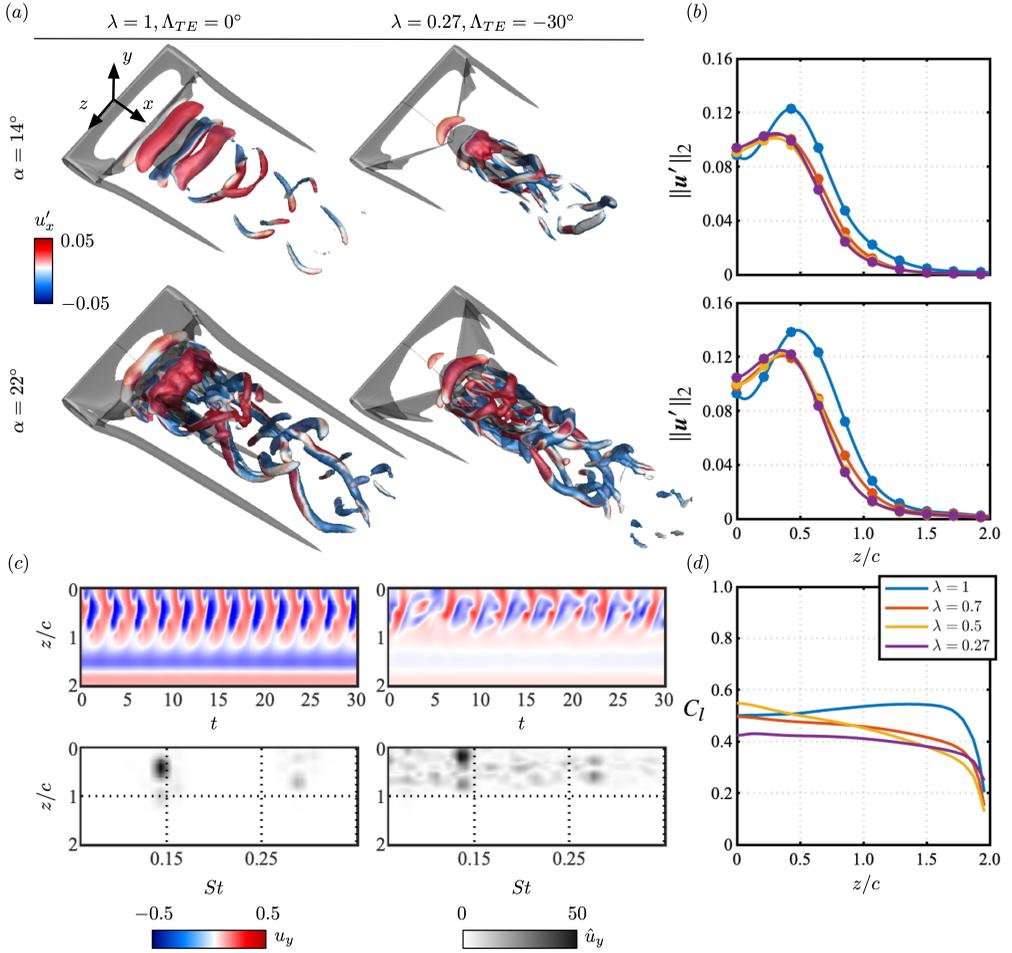
\begin{figure}
		\begin{tikzpicture}
		\node[anchor=south west,inner sep=0] (image) at (0,0) {\includegraphics[trim=0mm 0mm 0mm 0mm, clip,width=1\textwidth]{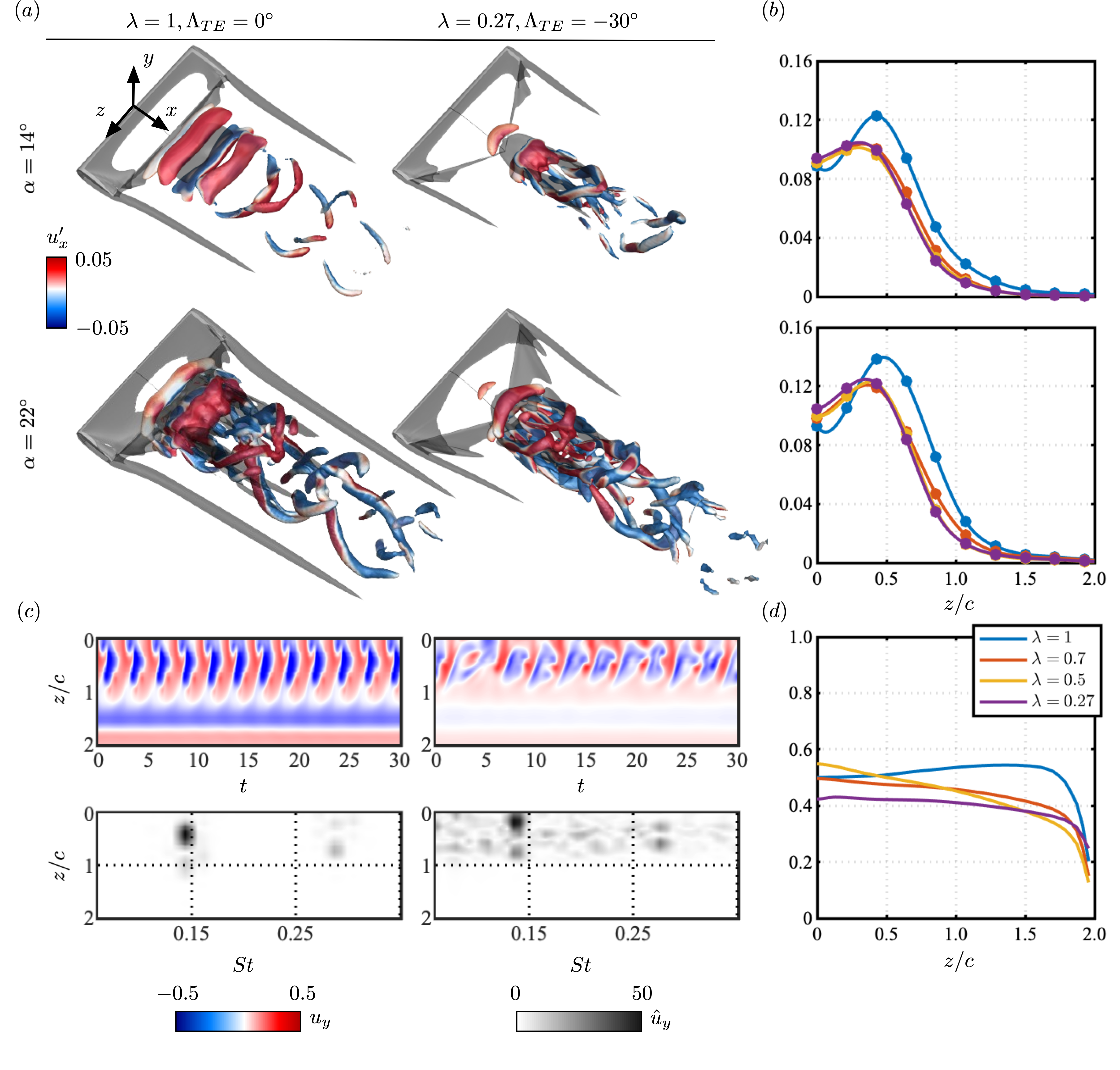}};
		\footnotesize
		\node[align=left] at (9.1,10.7) {\rotatebox{90}{$\|\mathbi{u}^\prime\|_2$}};
		\node[align=left] at (9.1,7.5) {\rotatebox{90}{$\|\mathbi{u}^\prime\|_2$}};
		\node[align=left] at (9.3,3.6) {\rotatebox{00}{$C_l$}};
		\end{tikzpicture} \vspace{-4mm}
		\caption{($a$) Isosurfaces of flow fields around tapered wings with $sAR = 2$, $\Lambda_\text{LE} = 0^\circ$, $\lambda = 0.27$ and $1$, $\alpha = 14^\circ$ and $22^\circ$. Time-averaged $\overline{Q} = 1$ isosurface is shown in gray. Instantaneous $Q^\prime = 0.2$ isosurface is shown colored by $u_x^\prime$. $(b)$ Spanwise distribution of $\| \mathbi{u}^\prime \|_2$ for different $\lambda$ for $\Lambda_\text{LE} = 0^\circ$ wings. $(c)$ Spatial-temporal (top) and PSD (bottom) of ${u}_y$ distribution over the spanwise direction from probes located at $(x,y)/c = (3,-0.5)$ for the $\lambda = 0.27$ and $1$ tapered wings at $\alpha = 22^\circ$ shown above. $(d)$ Sectional lift distribution over wingspan for tapered wings at $\alpha = 22^\circ$. } 
		\label{fig:qcritOverview_TE}
	\end{figure}
	
	Let us take a closer look at the effect of wing taper for unswept LE wings with forward-swept TE, as it allows us to isolate the $\Lambda_\text{TE}$ effect on the wake dynamics. For tapered wings with $\lambda = 0.27, 0.5, 0.7$, and $1$, the planforms we study in this section have  $\Lambda_\text{TE} = -30^\circ, -18.4^\circ, -10^\circ$, and $0^\circ$, respectively. The negative $\Lambda_\text{TE}$ indicates forward sweep. The LE is fixed with $\Lambda_\text{LE} = 0^\circ$. For such wings, taper has a negative impact on the aerodynamic performance, while concentrating the unsteady shedding to a narrow region near the root, and significantly reducing the tip vortex strength, as shown in figure \ref{fig:qcritOverview_TE}($a$). 
	
\begin{figure}
		\begin{tikzpicture}
		\node[anchor=south west,inner sep=0] (image) at (0,0) {\includegraphics[trim=0mm 0mm 0mm 0mm, clip,width=1\textwidth]{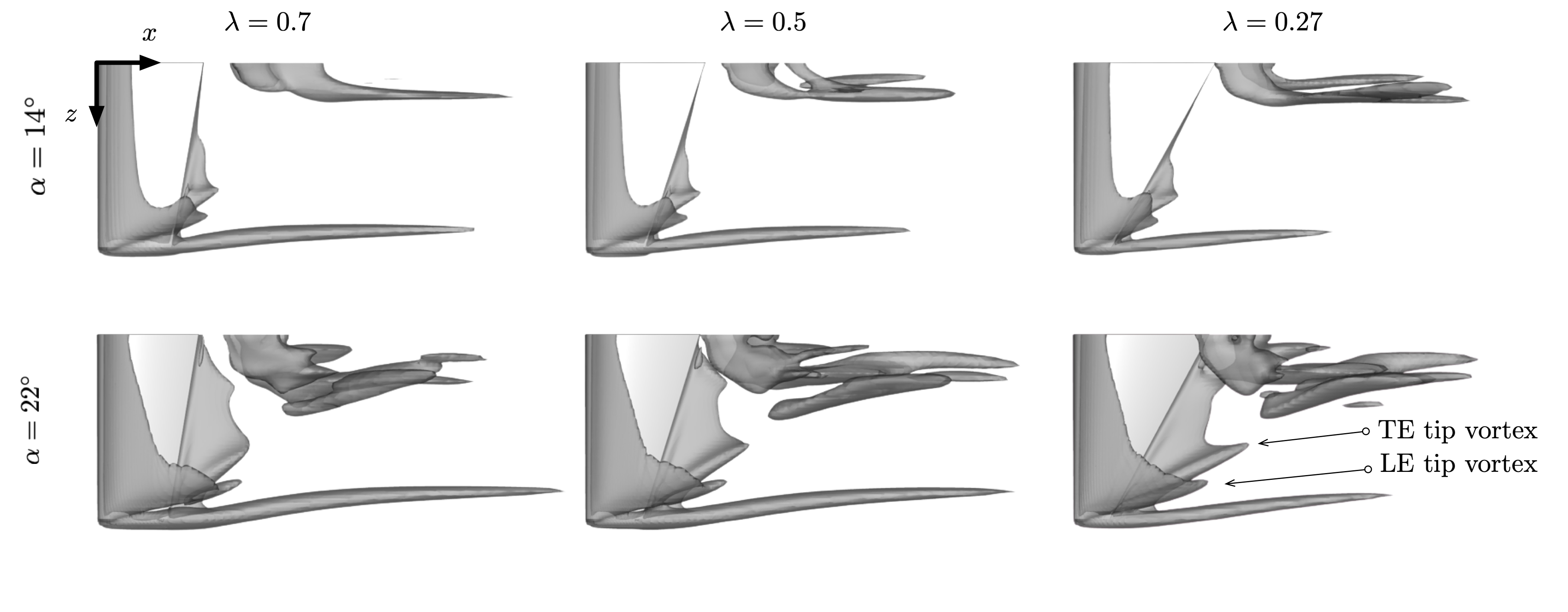}};
		\footnotesize
		\end{tikzpicture} \vspace{-10mm}
		\caption{Vortical structures emerging near the wing tip around tapered wings with unswept LE and forward-swept TE. Top view of time-averaged flow fields around  $sAR = 2$ tapered wings at $\alpha = 14^\circ$ and $22^\circ$, with $\Lambda_\text{LE} = 0^\circ$, visualized using gray-colored time-averaged isosurfaces of $\overline{Q} = 2$.} 
		\label{fig:secondaryTipVortex}
	\end{figure}

	Tapered wings have a smaller $c_\text{tip}$, which weakens the tip vortices and decreases its length, alleviating the inboard downwash over the wing. Such tip vortex attenuation and the aforementioned concentration of shedding over the root region occur for all angles of attack shown herein. The influence of the incidence angle appears on the formation of secondary vortices near the wing tip. For wings at high incidence angle, a secondary tip vortex is known to emerge from the LE, as shown  in figure \ref{fig:secondaryTipVortex} \citep{Devoria:JFM17,Zhang:JFM20}. For the tapered wings with forward-swept TE at $\alpha = 22^\circ$, there is also another core vortex that emerges near the wing tip from the TE. This structure is seen over the vortex sheet rolling up the TE as a slanted vortex pointing toward the root for the lower taper ratio, as visualized in figure \ref{fig:secondaryTipVortex}.
	
	To gain further insights on the characteristics of wake vortices, we study the unsteady flow behavior over the wingspan using probe measurements of velocity fluctuations over $x/c \in [3,4]$, $y/c \in [-1.5,0.5]$. The $x/c$ location is arbitrary and does not affect significantly the results. The $y/c$ range encompasses the region where vortical structures appear. Over this region, we probe the norm of the root-mean-square (RMS) of the velocity, $\|\mathbi{u}^\prime\|_2$. This measurement represents the spanwise distribution of flow unsteadiness, as shown in figure \ref{fig:qcritOverview_TE}($b$).
	
	By examining at the spanwise $\|\mathbi{u}^\prime\|_2$ distribution in figure \ref{fig:qcritOverview_TE}($b$) for untapered wings (blue), we notice that the flow unsteadiness peaks at $z/c \approx 0.5$ and decays towards the wing tip for both angles of attack. For tapered wings, the spanwise $\|\mathbi{u}^\prime\|_2$ curves are independent of the taper ratio for $\lambda \le 0.7$.  For such wings, taper yields an attenuation of the $\|\mathbi{u}^\prime\|_2$ peak. The peak of $\|\mathbi{u}^\prime\|_2$ also moves towards $z/c \approx 0$,  showing a concentration of unsteadiness towards the wing~root for tapered wings.
	
	Next, we analyze the spatial-temporal distribution of $u_y$ from probes located at $(x,y)/c = (3,-0.5)$ over the spanwise direction, to investigate how wing taper affects the shedding behavior. Herein, temporal frequency is characterized through the Strouhal number defined as $St = f (c \sin \alpha / U_\infty)$,	where $f$ is the frequency. For comparison, the wake spectra for the flow over an untapered wing is shown on the left of figure \ref{fig:qcritOverview_TE}($c$). For this wing, there is a narrow peak of oscillations at $St \approx 0.14$. The wake spectra is clean with a vortex shedding pattern comprised of spanwise-dominated vorticity near the root, forming hairpin vortices and a steady streamwise vortex at the wing tip.	For the tapered wing, the spectra is broadband as a result of the mixing of streamwise and spanwise vortices near the wing root. Even though the wake exhibits more mixing, the spanwise structures remain dominant, being related to the PSD peak at $St \approx 0.13$. We note that the PSD peak occurs at a lower $St$ than the one observed for the untapered wing, as the core unsteady structures that populate the downstream wake arise from the root region of the wing, where the chord-length~is~large.
	
	As the post-stall wakes around tapered wings with forward-swept TE concentrate shedding near the root, they also alter the sectional load distribution as the near-wake vortices play an important role in generating lift and drag over the wing. While untapered wings exhibit a peak in sectional lift near the tip region, we note that the root contribution to lift is higher for tapered wings, as shown in figure \ref{fig:qcritOverview_TE}$(d)$. Such load distribution is generally positive for flight stability \citep{Anderson:10}. For laminar post-stall flows over wings, the emergence of near-wake vortices closer to the wing surface can provide added lift \citep{Lee:JFM12,Zhang:PRF22}. As shown in figure \ref{fig:qcritOverview_TE}$(a)$, there are fewer large near-wake structures over the wing for the lower $\lambda$.  This is a possible reason of the decrease in $C_l$ over the entire wingspan experienced by the $\lambda = 0.27$ wing.

	\subsubsection{Tapered wings with backward-swept LE and unswept TE}
	\label{sec:LEsweep}

	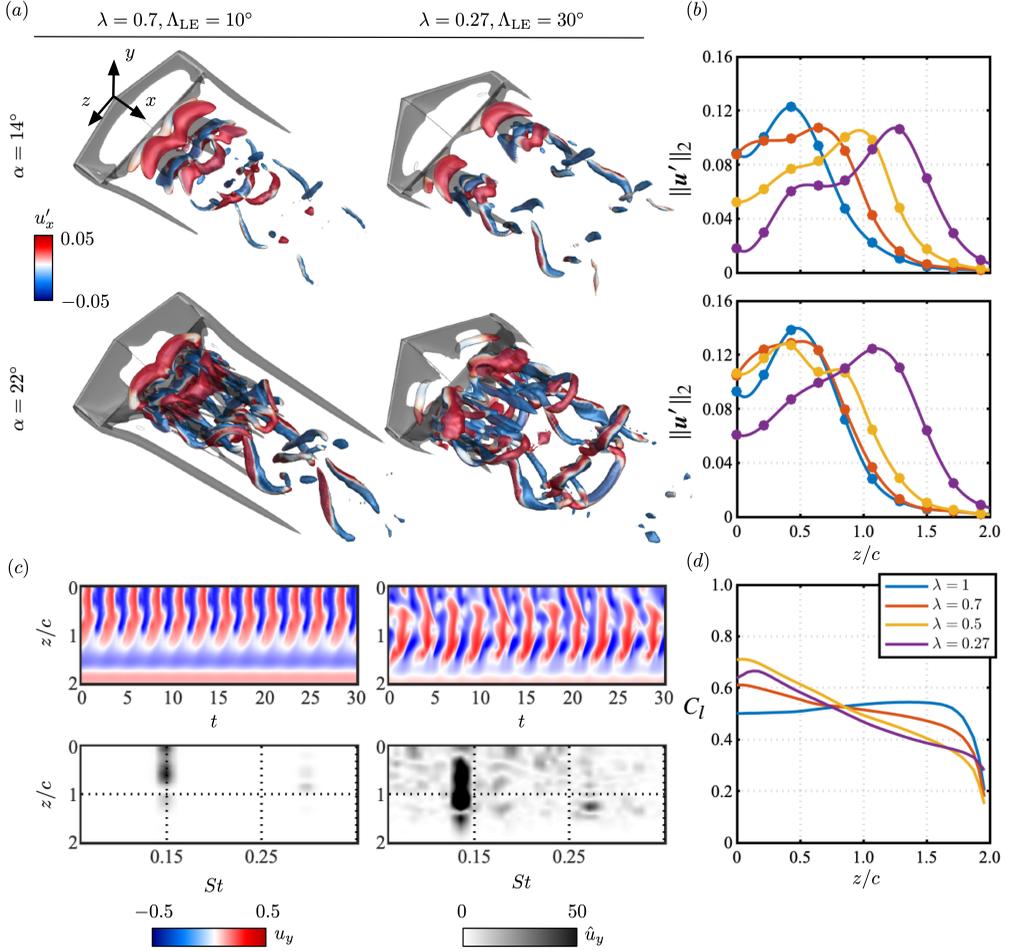
\begin{figure}
		\begin{tikzpicture}
		\node[anchor=south west,inner sep=0] (image) at (0,0) {\includegraphics[trim=0mm 0mm 0mm 0mm, clip,width=1\textwidth]{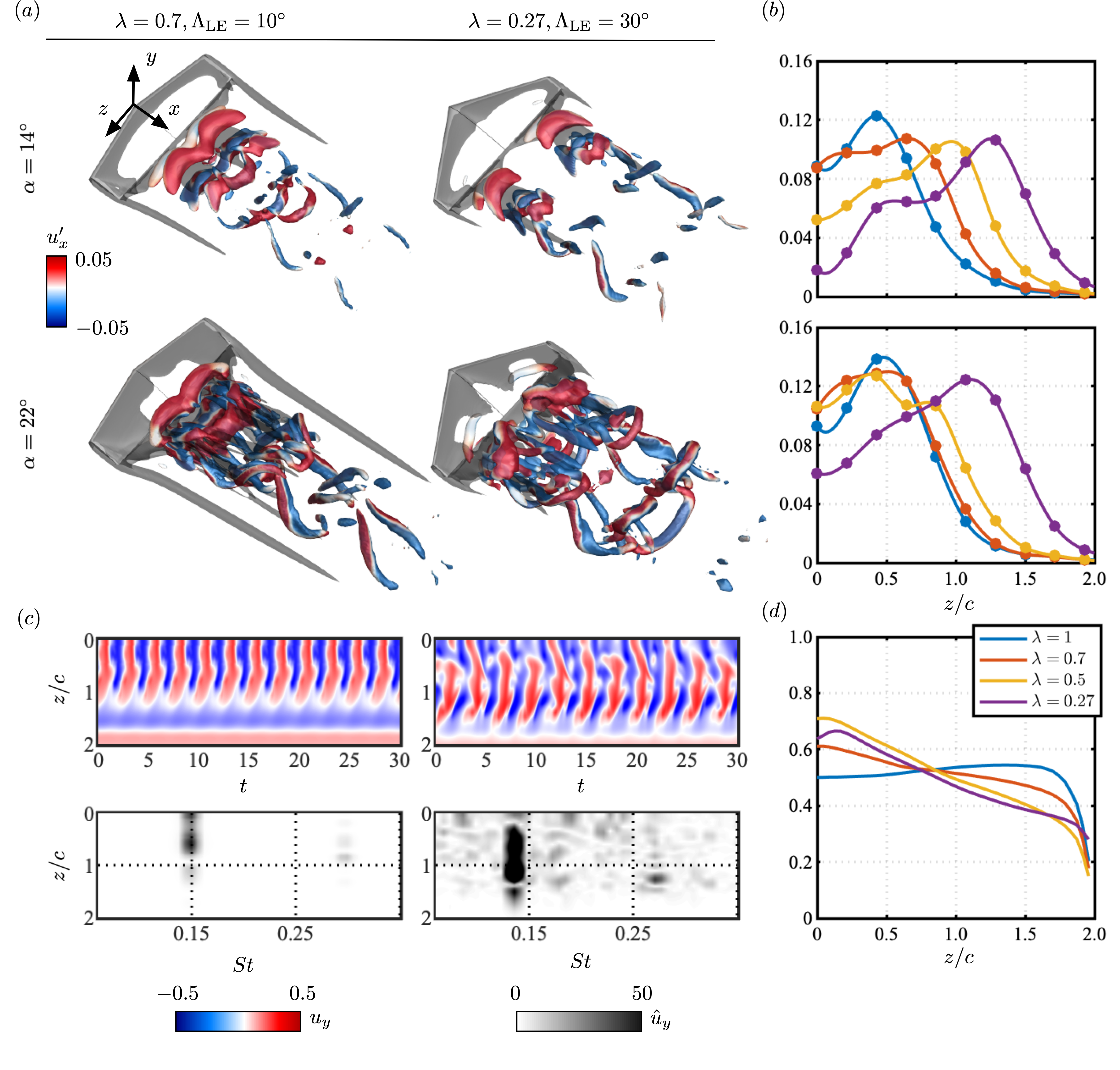}};
		\footnotesize
		\node[align=left] at (9.1,10.7) {\rotatebox{90}{$\|\mathbi{u}^\prime\|_2$}};
		\node[align=left] at (9.1,7.5) {\rotatebox{90}{$\|\mathbi{u}^\prime\|_2$}};
		\node[align=left] at (9.3,3.6) {\rotatebox{00}{$C_l$}};
		\end{tikzpicture} \vspace{-4mm}
		\caption{Isosurfaces of flow fields around tapered wings with $sAR = 2$, $\Lambda_\text{TE} = 0^\circ$, $\lambda = 0.27$ and $0.7$, $\alpha = 14^\circ$ and $22^\circ$. Time-averaged $\overline{Q} = 1$ isosurface is shown in gray. Instantaneous $Q^\prime = 0.2$ isosurface is shown colored by $u_x^\prime$. $(b)$ Spanwise distribution of $\| \mathbi{u}^\prime \|_2$ for different $\lambda$ for $\Lambda_\text{LE} = 0^\circ$ wings. $(c)$ Spatial-temporal (top) and PSD (bottom) of ${u}_y$ distribution over the spanwise direction from probes located at $(x,y)/c = (3,-0.5)$ for the $\lambda = 0.27$ and $0.7$ tapered wings at $\alpha = 22^\circ$ shown above. $(d)$ Sectional lift distribution over wingspan for tapered wings at $\alpha = 22^\circ$. } 
		\label{fig:qcritOverview_LE}
	\end{figure}
	
	Next, let us analyze the taper effects of wings with backward-swept LE and fixed unswept TE, to understand and separate the effects of the $\Lambda_\text{LE}$ on the global wake. For such wings with $\lambda = 0.27, 0.5, 0.7$, and $1$, the planforms have  $\Lambda_\text{LE} = 30^\circ, 18.4^\circ, 10^\circ$, and $0^\circ$, respectively. The positive $\Lambda_\text{LE}$ indicates backward sweep. The TE is fixed with $\Lambda_\text{TE} = 0^\circ$. For such wings, taper yields an opposite effect on the wake characteristics, when compared to those discussed in section \ref{sec:TEsweep}. As shown in figure \ref{fig:ClandCd}, such wake pattern results in a better aerodynamic performance for wings with the same $\lambda$ but distinct LE and TE sweep angles. Herein, taper shifts the unsteadiness region toward the wing tip, as shown in figure \ref{fig:qcritOverview_LE}($a$).  
	
	Concurrently, the tip vortex weakens for tapered wings with the shortened $c_\text{tip}$, which alleviates of the inboard downwash near the tip, similar to what was observed for the wings in section \ref{sec:TEsweep}. This increases the effective angle of attack near the tip and allows for the flow to detach from the wing surface and form wake shedding structures near $z/c \approx 1$, as shown in figure \ref{fig:qcritOverview_LE}($a$). We quantify the effect of wing taper on flow unsteadiness through the winsgpan distribution of $\|\mathbi{u}^\prime\|_2$, as shown in figure \ref{fig:qcritOverview_LE}($b$). For both angles of attack, taper affects the wake shedding distribution over the wingspan. For $\lambda = 0.27$ (purple), at $\alpha = 22^\circ$, the peak of $\|\mathbi{u}^\prime\|_2$ appears near the quarter-span at $z/c \approx 1.25$, with a gradual transition towards $z/c \approx 0.5$ from $\lambda =0.27$ to $1$.
	
	As seen in figure \ref{fig:qcritOverview_LE}($b$), tapered wings with backward-swept LE and unswept TE exhibit unsteadiness over a larger spanwise length than untapered wings. For instance, let us observe the spanwise $\|\mathbi{u}^\prime\|_2$ distribution for wings at $\alpha = 22^\circ$. For the untapered wing (blue), $\mathbi{u}^\prime \ge 0.02$ over $0 \le z/c \le 1$, which is the region where significant unsteady wake structures appear. Now, for the tapered wing with $\lambda = 0.27$, $\mathbi{u}^\prime \ge 0.02$ over $0 \le z/c \le 1.6$, hence large unsteady structures can be observed over a larger spanwise portion of the wake. 
	
	The spatial-temporal distribution of the transverse velocity $u_y$ over the spanwise direction also shows that the wake of backward-swept LE and unswept TE tapered wings exhibits $3$-D vortical structures that result in a broadband wake spectrum, as shown in figure \ref{fig:qcritOverview_LE}($c$). The wake, however, is mainly dominated by large quasi-$2$-D spanwise aligned vortex rolls observed for all taper ratios. For $\lambda = 0.27$, as unsteadiness appears over a larger portion of the  wingspan, the stronger shedding structures are  hairpin-like vortices that appear between $0.5 \le z/c \le 1.5$, as shown on the right of figure \ref{fig:qcritOverview_LE}($c$). 
	
	\begin{figure}
		\begin{tikzpicture}
		\node[anchor=south west,inner sep=0] (image) at (0,0) {\includegraphics[trim=0mm 0mm 20mm 0mm, clip,width=1\textwidth]{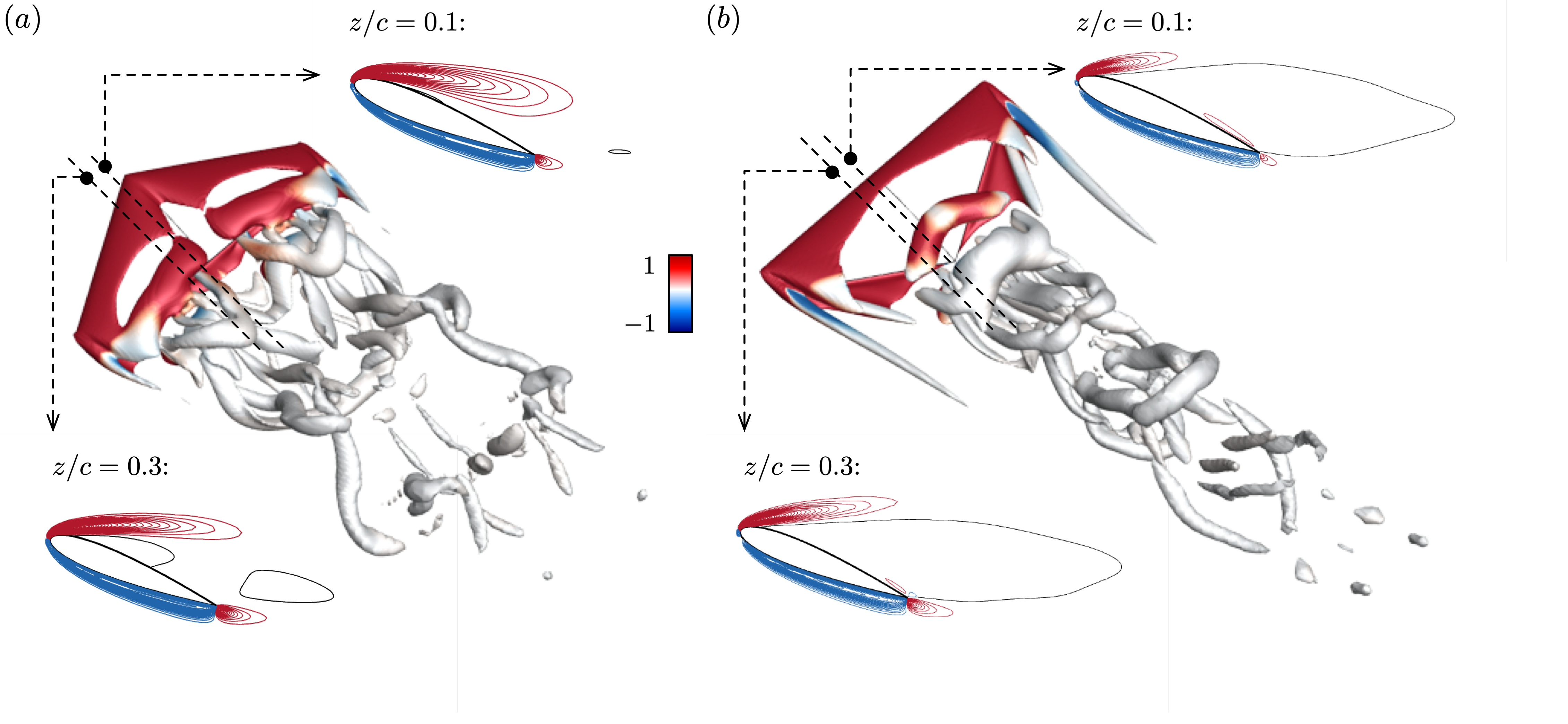}};
		\scriptsize
		\node[align=left] at (5.9,4.5) {\rotatebox{0}{$(\mathbf{u} \times \bs{\omega}) \cdot \nabla \phi_y$}};
		\end{tikzpicture} \vspace{-8mm}
		\caption{Lift elements around $\lambda = 0.27$ tapered wings at $\alpha = 22^\circ$ at the time of maximum lift. $(a)$ backward-swept LE with unswept TE and $(b)$ Unswept LE with forward-swept TE . Perspective view shown with isosurfaces of $Q = 1$ colored by $(\mathbf{u} \times \bs{\omega}) \cdot \nabla \phi_y$. Slices at selected $z/c$ locations with isocontours of lift elements and a black solid line over $\overline{u_x} = 0$.} 
		\label{fig:ForceElements}
	\end{figure}
	
	The sectional lift distribution, shown in figure \ref{fig:qcritOverview_LE}$(d)$, reveal that tapered wings with backward-swept LE and unswept TE significantly increase the root contribution and reduce the influence of the near tip  region on the overall lift. The increase in root contribution to lift results from the shifting of the separation bubble toward the tip. This shifting causes a pair of near-wake vortices to emerge over the wing surface at the root region, as shown in figure \ref{fig:qcritOverview_LE}$(a)$. Using force element analysis \citep{Chang:PRSA92}, we reveal the near-wake structures that contribute to lift. In this approach, the volume force elements  are identified by the dot product of the Lamb vector $\mathbf{u} \times \bs{\omega}$ and an auxiliary potential $\nabla \phi_i$ (details in appendix \ref{sec:forceleements}). Force element analysis shows a vortex pair emerging near the root, which increases the local contribution to the total lift over the wing.
	
	In figure \ref{fig:ForceElements}, force elements further show that vortical structures with major contribution to lift appear over the separation bubble. Here, this region is illustrated by a black solid line contour at $\overline{u_x} = 0$ on the $2$-D slices at $z/c = 0.1$ and $0.3$. In particular, the emergence of the near-root vortex pair is persistent for wings with backward-swept LE as similar structures have been identified for backward-swept untapered wings by \cite{Zhang:PRF22}. These structures are absent for unswept LE wings both tapered and untapered, as shown in figure \ref{fig:ForceElements}$(b)$. In fact, force elements over tapered unswept LE wings show that the lift elements emerging over the wing are much smaller than the ones over tapered wings with backward-swept LE. As the separation bubble near the root becomes larger over tapered wings with unswept LE, the wake structures are shifted far from  the wing, reducing their contribution to the total lift.
	
	\begin{figure}
		\begin{tikzpicture}
		\node[anchor=south west,inner sep=0] (image) at (0.0,0.0) {\includegraphics[trim=0mm 0mm 0mm 0mm, clip,width=1.0\textwidth]{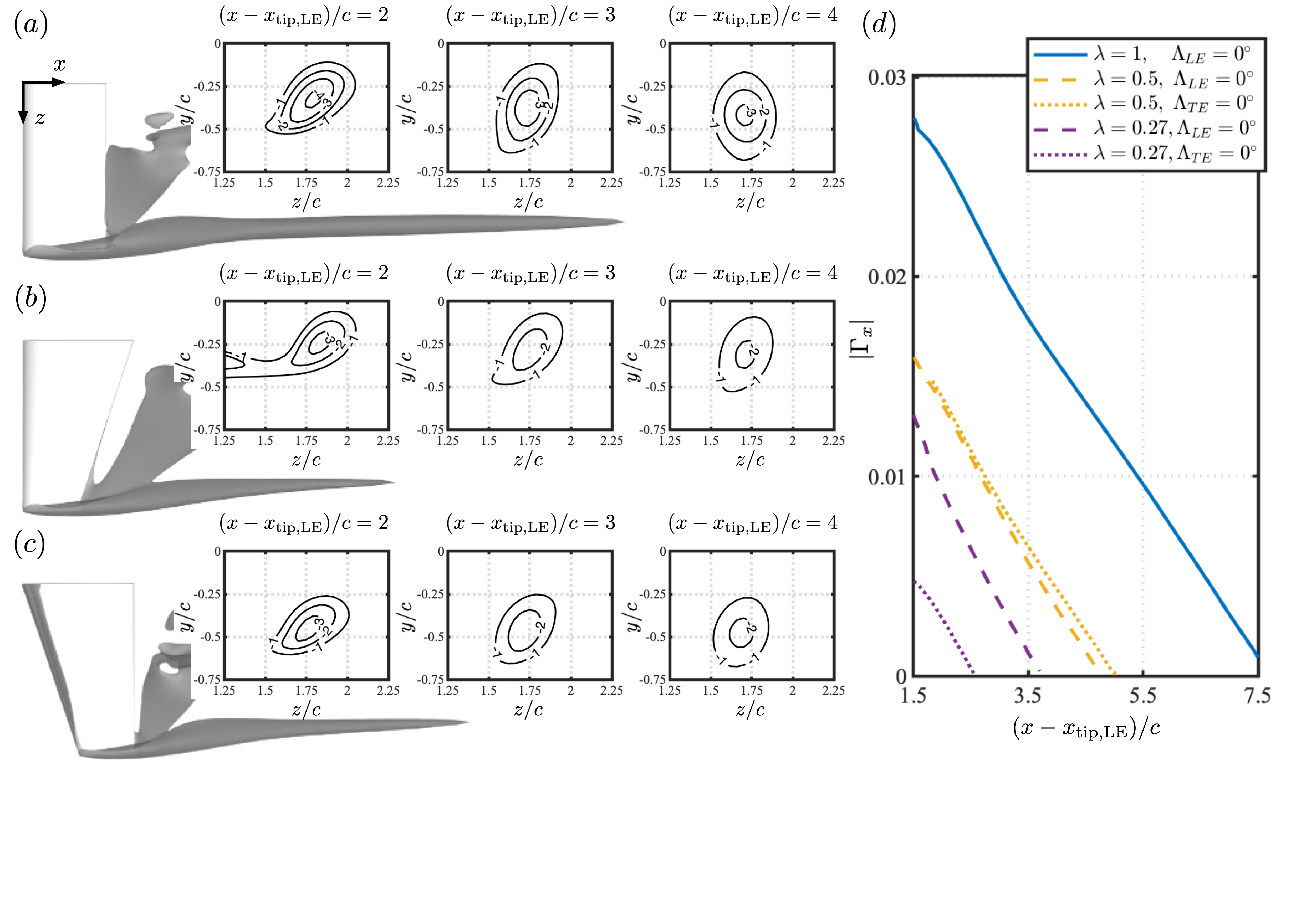}};
		\end{tikzpicture} \vspace{-18mm}
		\caption{Streamwise circulation $|\Gamma_x|$ of tip vortex around $(a)$ an untapered wing and tapered wings with $\lambda = 0.5$ with $(b)$ unswept LE and forward-swept TE and $(c)$ backward-swept LE and unswept LE at $\alpha = 22^\circ$. Flow field visualized with grey-colored isosurfaces of $\overline{\omega}_x = -2$ and $2$-D slices with isolines of $\overline{\omega}_x$ at specific $(x - x_\text{tip,LE})/c$ locations. The magnitude of $|\Gamma_x|$ computed for the isocontour of $(d)$ $\overline{\omega_x} = -2$ for tapered swept wings with different~planform~configurations.} 
		\label{fig:tipVortex}
	\end{figure}

	The tip vortex is also substantially affected by wing taper, becoming smaller than the tip vortex around untapered wings, as shown in figure \ref{fig:tipVortex}($a$). Tip vortices have high importance in terms of the aerodynamic characteristics of the wing \citep{Francis:JA79,Green:JFM91,Devenport:JFM96,Birch:JA04,Taira:JFM09,Zhang:JFM20,Dong:EF20,Toppings:JFM21,Toppings:JFM22} and, in the case of tapered wings, due to the small $c_\text{tip}$, tip vortices are attenuated, as a result of the reduced pressure differences between upper and lower side of the wing near the tip. Beyond that, even for wings with the same $\lambda$, the tip vortex behavior can be shifted in $x$-direction depending on how the wing is tapered, whether it has a backward-swept LE or a forward-swept TE, as shown in figure \ref{fig:tipVortex}($b$,$c$). For this reason, we must also analyze the tip vortex at a distance $x/c$ from the LE at the wing tip, which is identified herein as $x_\text{tip,LE}$.
	
	We can observe how wing taper affects the strength of the tip vortex by analyzing the $\overline{\omega_x}$ near the tip, as shown in figure \ref{fig:tipVortex}($a$-$c$). The isosurfaces of $\overline{\omega_x}$ and the contour lines at representative $(x - x_\text{tip,LE})/c$ locations show the decay of vorticity magnitude for tapered wings with $\lambda = 0.5$. However, the effect of taper is not the same for both wings, even though they share the same taper ratio. This difference can be quantified as we compute the streamwise circulation $\Gamma = \int_{C} = \mathbi{u} \cdot \text{d}\mathbi{l}$. Here, $C$ is the isocontour of $\overline{\omega_x} = -2$, as shown in figure \ref{fig:tipVortex}($d$). The choice of $\overline{\omega_x}$ level is carefully chosen to isolate the tip vortex.  
	
	The tip vortex diffuses downstream of the wing, which makes the $|\Gamma_x|$ profiles decay slowly \citep{Edstrand:JFM18b,Zhang:JFM20}. In general, for tapered wings the reduction in $c_\text{tip}$ is the main cause of the tip vortex weakening, thus the $|\Gamma_x|$ circulation decays with $\lambda$ at any distance from the wing tip. The circulation $|\Gamma_x |$ further reveals how different types of wing taper can affect the strength of the tip vortex, as shown in figure \ref{fig:tipVortex}($d$). At any given streamwise distance from the wing tip at the LE, $(x - x_\text{tip,LE})/c$, for $\lambda = 0.5$, the streamwise circulation decay is similar for both tapered wings. For $\lambda = 0.27$, the tip vortex strength decays considerably depending on the LE and TE sweep angles. For tapered wings with backward-swept LE and unswept TE, the root shedding is shifted toward the tip region and reduces the tip vortex strength at any distance from the TE, when compared to the wing with same $\lambda$ and forward-swept TE. For forward-swept TE wings, as vortex shedding is concentrated near the wing root, it has a minor influence on the tip vortex.
	
	\subsubsection{Tapered wings with high LE sweep angles}
	\label{sec:SWandTaper}
	
	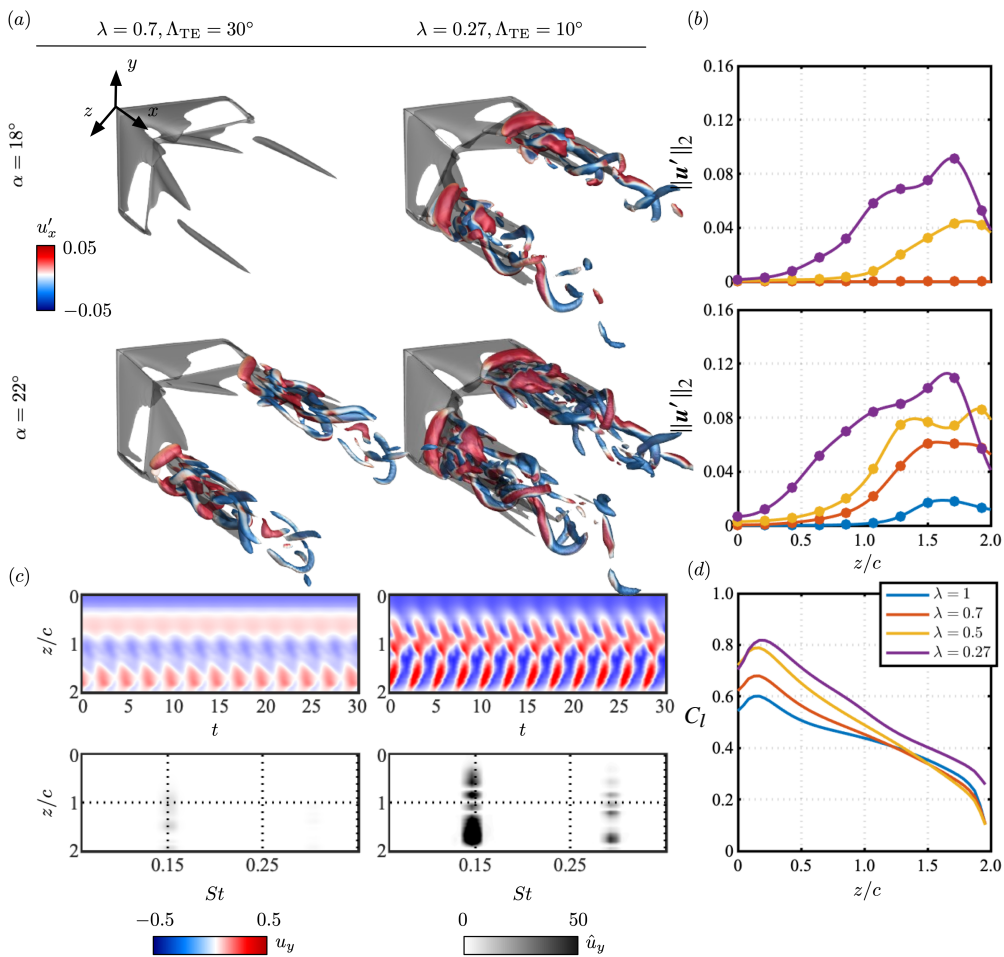
\begin{figure}
		\begin{tikzpicture}
		\node[anchor=south west,inner sep=0] (image) at (0,0) {\includegraphics[trim=0mm 0mm 0mm 0mm, clip,width=1\textwidth]{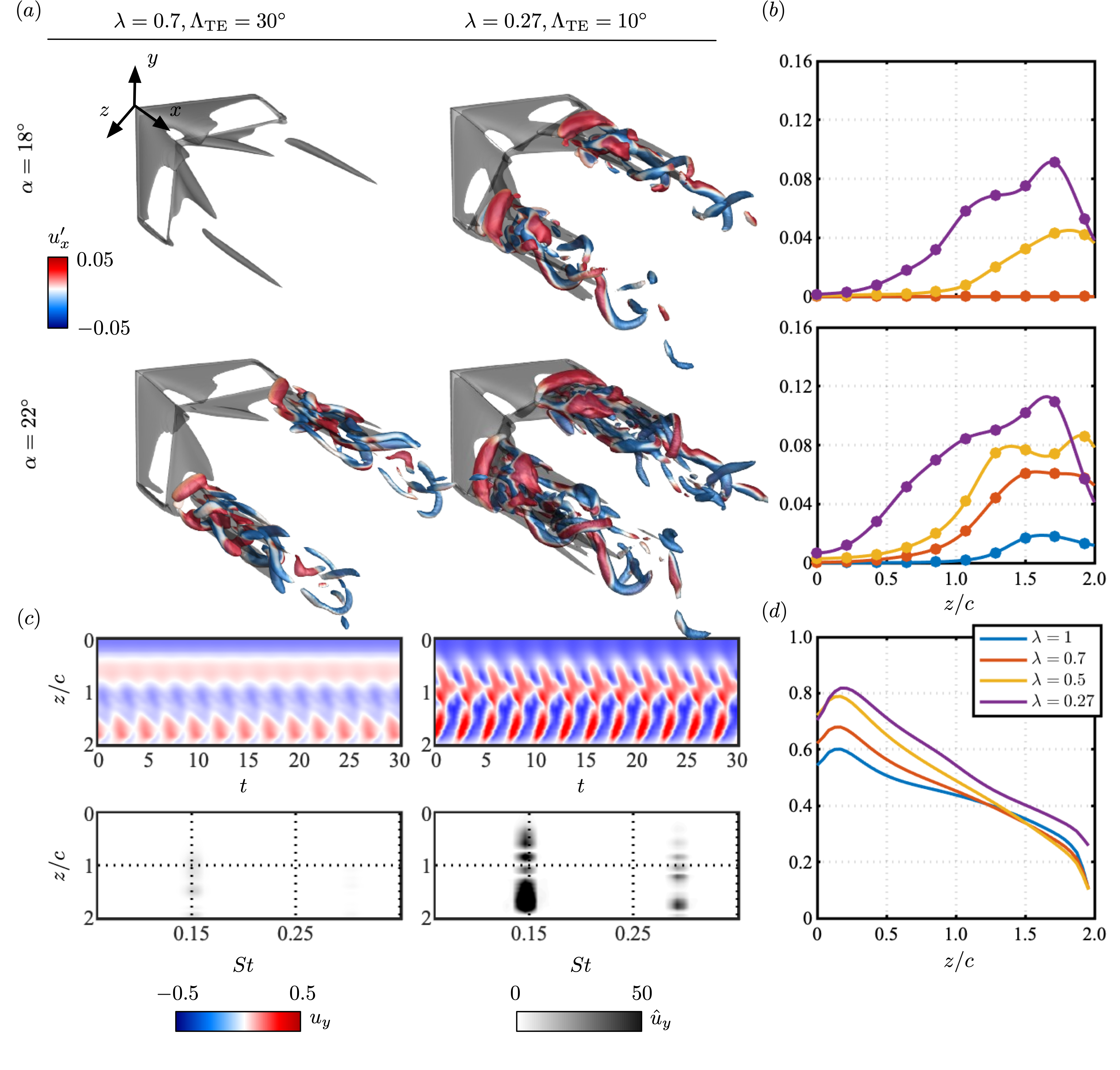}};
		\footnotesize
		\node[align=left] at (9.1,10.9) {\rotatebox{90}{$\|\mathbi{u}^\prime\|_2$}};
		\node[align=left] at (9.1,7.7) {\rotatebox{90}{$\|\mathbi{u}^\prime\|_2$}};
		\node[align=left] at (9.3,3.6) {\rotatebox{00}{$C_l$}};
		\end{tikzpicture} \vspace{-4mm}
		\caption{Isosurfaces of flow fields around tapered swept wings with $sAR = 2$, $\Lambda_\text{LE} = 40^\circ$, $\lambda = 0.27$ and $0.7$, $\alpha = 18^\circ$ and $22^\circ$. Time-averaged $\overline{Q} = 1$ isosurface is shown in gray. Instantaneous $Q^\prime = 0.2$ isosurface is shown colored by $u_x^\prime$. $(b)$ Spanwise distribution of $\| \mathbi{u}^\prime \|_2$ for different $\lambda$ for $\Lambda_\text{LE} = 0^\circ$ wings. $(c)$ Spatial-temporal (top) and PSD (bottom) of ${u}_y$ distribution over the spanwise direction from probes located at $(x,y)/c = (3,-0.5)$ for the $\lambda = 0.27$ and $0.7$ tapered wings at $\alpha = 22^\circ$ shown above. $(d)$ Sectional lift distribution over wingspan for tapered wings at $\alpha = 22^\circ$. } 
		\label{fig:qcritOverview_SW}
	\end{figure}
	
	The final class of tapered wing shapes we consider are those with high LE sweep. For the swept wings discussed herein, with $\Lambda_\text{LE} > 30^\circ$, wake oscillations are strongly attenuated. For laminar flows over untapered wings with high sweep angles at moderate angles of attack, the wake becomes steady, while at high angles of attack, unsteadiness may develop in the wing tip region \citep{Zhang:JFM20b,Ribeiro:JFM23triglobal}. For highly swept and tapered wings, the flow exhibits wake shedding for small $\lambda$, as shown in~figure~\ref{fig:qcritOverview_SW}($a$). 
	
	Here, we analyze wings with a fixed $\Lambda_\text{LE} = 40^\circ$, while the TE is swept with $\Lambda_\text{TE} = 10^\circ, 21.6^\circ, 30^\circ$, and $40^\circ$ for $\lambda = 0.27, 0.5, 0.7$, and $1$, respectively. The onset of shedding for highly swept and tapered wings results from the distinct effects of $\Lambda_\text{LE}$ and $\Lambda_\text{TE}$. For the present tapered swept wings, the vortical structures emerging from the TE promote unsteadiness in the wake near the wing tip. For lower taper ratios, wings have a low $\Lambda_\text{TE}$, which causes wake oscillation to appear and become large towards the root. Such effects show that while the high $\Lambda_\text{LE}$ has the effect of stabilizing wake oscillations for untapered wings,  the combination of wing taper and sweep can promote wake unsteadiness.  
	
	For instance, at $\alpha = 18^\circ$ the wake is steady for $\lambda = 0.7$ with long steady streamwise vortices developing from both LE and TE. At $\lambda = 0.5$, unsteadiness appears with vortex rolls at the wing tip, with wake shedding appearing for $\lambda = 0.27$. We quantify the wing taper effect in figure \ref{fig:qcritOverview_SW}($b$). For instance, for the wings with $\lambda \ge 0.7$ at $\alpha = 18^\circ$, the flow is steady and $\|\mathbi{u}^\prime\|_2$ is negligible in the wake. At $\alpha = 22^\circ$, $\|\mathbi{u}^\prime\|_2$ is small for untapered wings, increasing considerably in magnitude and spanwise length as the taper ratio decreases. For highly swept tapered wings, the flow fluctuations are exhibited at the tip, further appearing over the midspan for the lower taper ratios.
	
	The unsteady vortices exhibited in the wakes of tapered wings with high LE sweep angles behave as vortex shedding structures, as shown by the probed $u_y$ in the wake in figure \ref{fig:qcritOverview_SW}($c$). For $\lambda = 0.7$, the vortices appear as a consistent flow oscillation departing from the wing tip. For $\lambda = 0.27$, the wake is dominated by spanwise-aligned roll structures that occupy a large portion of the wingspan. As these structures develop from the wingspan region near the wing tip, that has a reduced chord length, their frequency $St \approx 0.15$ is slightly higher than the shedding frequency of untapered wings. 
	
	The sectional lift distributions around tapered wings with high LE sweep, seen in figure \ref{fig:qcritOverview_LE}$(d)$,  show how wing taper substantially increases the root contribution to the overall lift. Backward swept wings have a higher contribution of lift from the root region due to the pair of vortical structures near the midspan that attach closer to the wing surface \citep{Zhang:PRF22} similar to the structures shown in figure \ref{fig:ForceElements}$(a)$. For tapered wings, the combined effect of the midspan vortex pair and the development of shedding structures near the wing results in a considerable increase of the overall lift over the entire wingspan for the $\lambda = 0.27$ wing.
	
	
	
	\section{Conclusions}
	\label{sec:conclusions}
	We have examined the influence of taper and sweep on the dynamics of wake structures for finite NACA 0015 wings with straight-cut tip at a Reynolds number of $600$ and a Mach number $0.1$. For this study, we performed an extensive campaign of direct numerical simulations of flows over half-span wings with symmetry boundary condition imposed at the wing root. The present numerical study spans over a wide parameter space with angles of attack between $14^\circ \le \alpha \le 22^\circ$, aspect ratios $sAR = 1$ and $2$, leading edge sweep angles $0^\circ \le \Lambda_\text{LE} \le 50^\circ$, and taper ratios between $0.27 \le \lambda \le 1$. This parameter space was chosen to characterize the effects of wing taper as well as the LE and TE sweep angles on the wake dynamics.
	
	Through direct numerical simulations, we observe that the flow over unswept and untapered wings forms a strong tip vortex, which interacts with the spanwise vortex detaching from the wing surface at the root region. This flow yields a three-dimensional and unsteady wake for all angles of attack considered herein. Untapered and swept wings are observed to advect the shedding region towards the wing tip for lower angles of sweep. At higher sweep angles, the wake oscillations are attenuated, yielding a steady wake around wings at lower angles of attack.
	
	Wing taper has a strong influence on the wake dynamics. For tapered wings, the LE and TE are not parallel and have a distinct influence on the flow structures within the stalled flow region.	For tapered wings with unswept LE and forward-swept TE, taper concentrates shedding structures towards the wing root and yields a broadband spectral content downstream in the wake as a result of increased mixing in that region. Beyond the unsteady wake shedding, the tip vortex is heavily affected by wing taper, reducing its length considerably for tapered wings, as the chord-length decreases towards the tip. 
	
	For tapered wings with backward-swept LE and unswept TE, the spanwise length where wake unsteadiness is observed increases as shedding is promoted over a larger portion of the wingspan. For this type of tapered wing planform, in contrast to the forward-swept wing effect, the peak of wake unsteadiness moves towards the wing tip region for lower taper ratios. Moreover, for wings with high LE sweep, although the flow is steady for $\lambda = 1$, taper causes wake unsteadiness to appear. The wake oscillations develop near the wing tip for moderate taper ratios. For low $\lambda$, wings with high LE sweep angles exhibit strong wake shedding structures occupying a large portion of the wingspan. 
	
	Through the detailed analysis of the wake structures, we also provide a map that classifies the wake behavior of tapered wings associating its behavior with the wing planform geometry and its angle of attack. The map provides an unique description of the overall flow physics of the wakes around tapered wings and reveals, for each semi aspect ratio and angle of attack, how the steady-unsteady flow behavior is related to the wing taper and LE sweep angle. The present study shows the effect of taper, as well as the effects of LE and TE sweep and evaluates its impact on the formation of the wake structures. 
	
	Lastly, we show how the wing taper affects the aerodynamic forces over the wing. We show that wings with the same taper ratio may present distinct overall lift and aerodynamic performance, as these characteristics are also influenced by the LE sweep of the wing. Our findings show that the combination of wing taper and high LE sweep can considerably improve lift and the aerodynamic performance of the wing in laminar post-stall flows conditions. The present insights gained on the effect of wing taper in the absence of turbulence serve as a stepping stone for future efforts that aim to study, interpret and control higher Reynolds number post-stall flows over tapered wings.
	
	\appendix
	\section{Grid verification}
	\label{sec:verification}
	
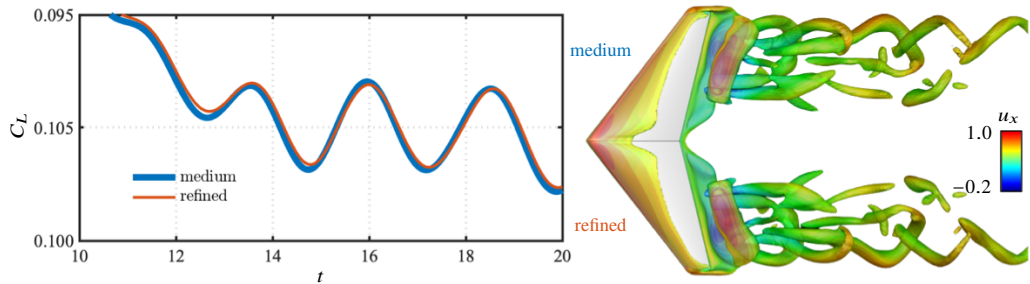
\begin{figure}
		\centering
		\begin{tikzpicture}
		\node[anchor=south west,inner sep=0] (image) at (0,0) {\includegraphics[trim=0mm 0mm 0mm 0mm, clip,width=1\textwidth]{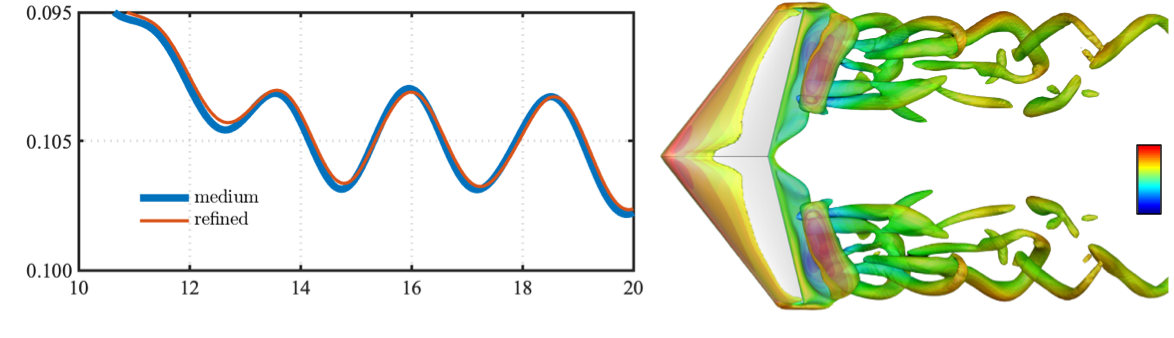}};
		\scriptsize
		\node[align=left] at (4.1,0.3) {\rotatebox{0}{$t$}};
		\node[align=left] at (0.1,2.3) {\rotatebox{90}{$C_L$}};
		\node[align=left] at (7.8,3.3) {\rotatebox{0}{\color{matblue}medium}};
		\node[align=left] at (7.8,1.0) {\rotatebox{0}{\color{matred}refined}};
		\node[align=left] at (13.2,2.4) {\rotatebox{0}{$u_x$}};
		\node[align=left] at (12.8,2.2) {\rotatebox{0}{$1.0$}};
		\node[align=left] at (12.7,1.5) {\rotatebox{0}{$-0.2$}};
		\end{tikzpicture} \vspace{-6mm}
		\caption{Lift coefficient over time and instantaneous flow field structures visualized using $Q = 1$ colored by instantaneous velocity component $u_x$ at the lift peak for the two sets of meshes used for grid verification for the wing with $(sAR,\alpha,\Lambda_\text{LE},\lambda) = (2,22^\circ,40^\circ,0.27)$.} 
		\label{fig:verification}
	\end{figure}

	We verify the convergence of grid resolution for the numerical results using a wing with $(sAR,\alpha,\Lambda_\text{LE},\lambda) = (2,22^\circ,40^\circ,0.27)$. This planform combines a high leading-edge sweep angle and the lowest taper ratio considered in the present study. Herein, we report the aerodynamic forces through their lift coefficients	$C_L$.  Two meshes are used for verification: a medium and a refined mesh. The medium mesh refinement is the one used throughout the present work. This mesh has $80$ grid points on both pressure and suction sides of the wing and $48$ grid points along the wingspan, with a total of approximately $3.1 \times 10^6$ control volumes. The refined mesh has $120$ grid points on pressure and suction sides, with $64$ grid points along the wingspan, resulting in approximately $4.3 \times 10^6$ control volumes in total. For the refined mesh we have increased the temporal resolution by setting the CFL to $0.5$. The quality of our medium mesh is assessed through the forces exerted over the wing and the instantaneous vortical elements as shown in figure \ref{fig:verification}. 
	
	\section{A portfolio of flow fields around tapered wings}
	\label{sec:flowviz}
\begin{figure}
		\centering
		\begin{tikzpicture}
		\node[anchor=south west,inner sep=0] (image) at (0,0) {\includegraphics[trim=0mm 0mm 0mm 0mm, clip,width=1\textwidth]{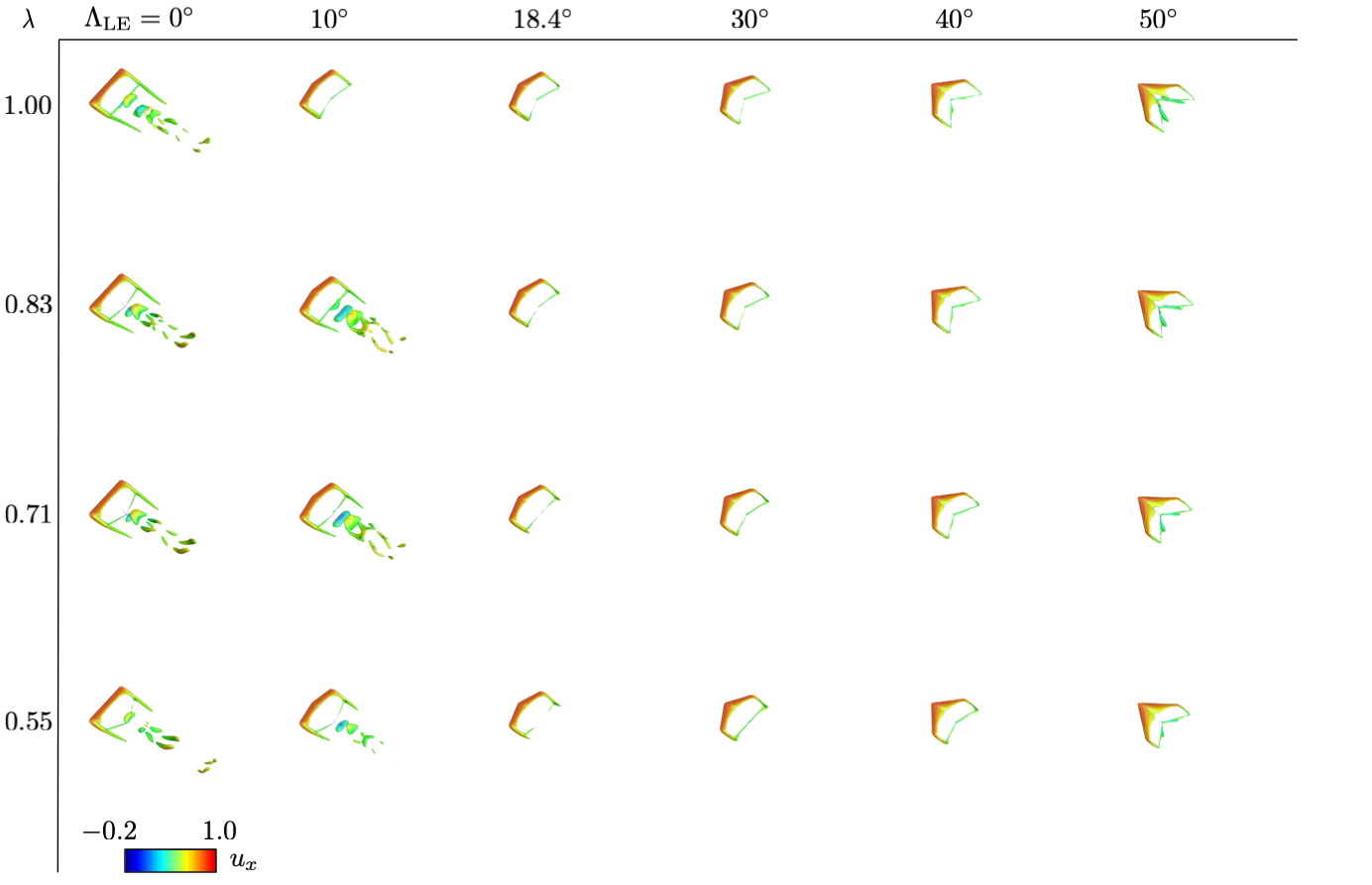}};
		\end{tikzpicture} \vspace{0mm}
		\caption{Instantaneous flow fields around wings of $sAR = 1$ at $\alpha = 14^\circ$ visualized using isosurfaces of $Q = 1$ colored by streamwise velocity $u_x$.} 
		\label{fig:perspective_ar1aoa14}
	\end{figure}

\begin{figure}
		\centering
		\begin{tikzpicture}
		\node[anchor=south west,inner sep=0] (image) at (0,0) {\includegraphics[trim=0mm 0mm 0mm 0mm, clip,width=1\textwidth]{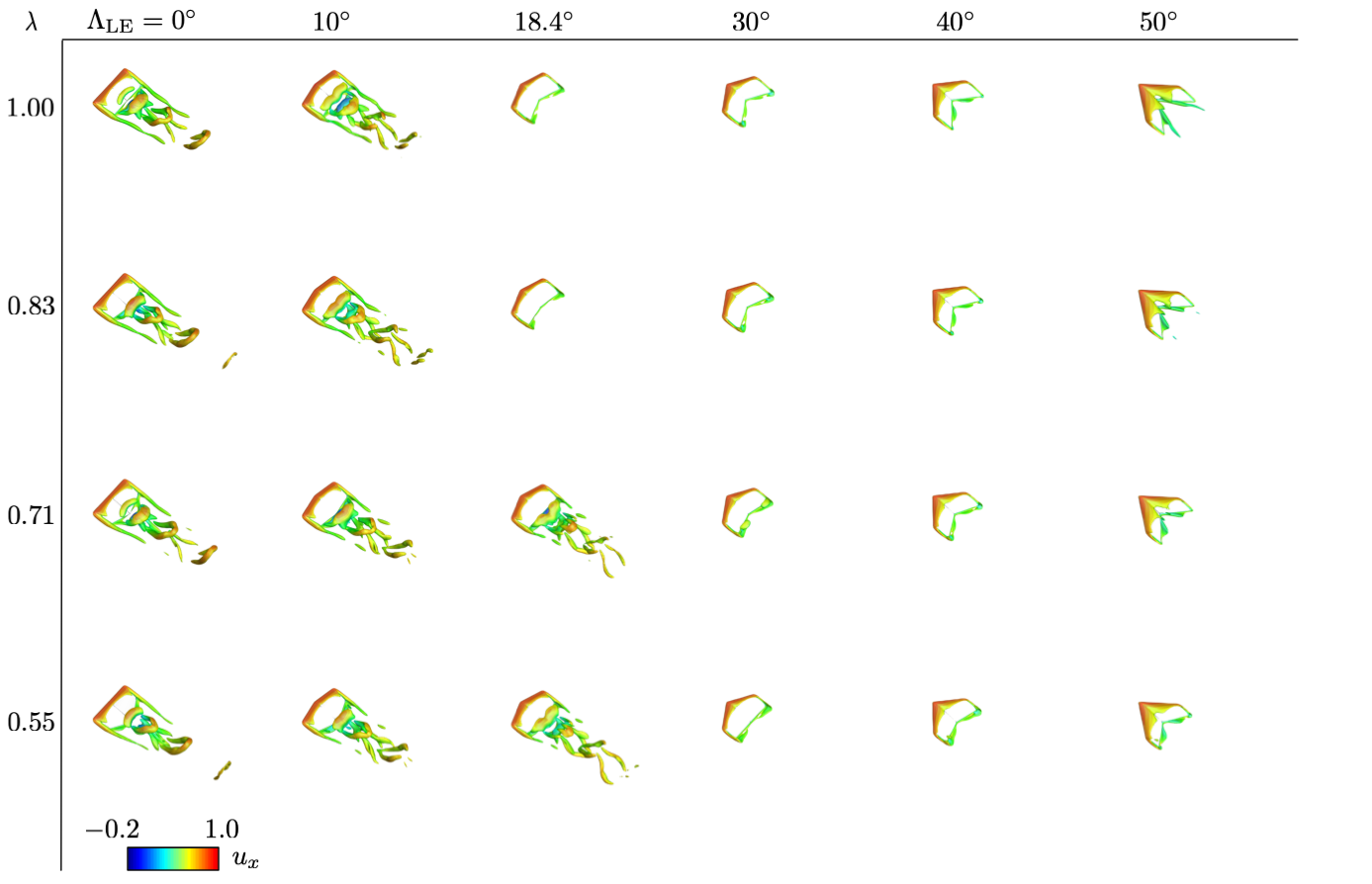}};
		\end{tikzpicture} \vspace{0mm}
		\caption{Instantaneous flow fields around wings of $sAR = 1$ at $\alpha = 18^\circ$ visualized using isosurfaces of $Q = 1$ colored by streamwise velocity $u_x$.} 
		\label{fig:perspective_ar1aoa18}
	\end{figure}

\begin{figure}
		\centering
		\begin{tikzpicture}
		\node[anchor=south west,inner sep=0] (image) at (0,0) {\includegraphics[trim=0mm 0mm 0mm 0mm, clip,width=1\textwidth]{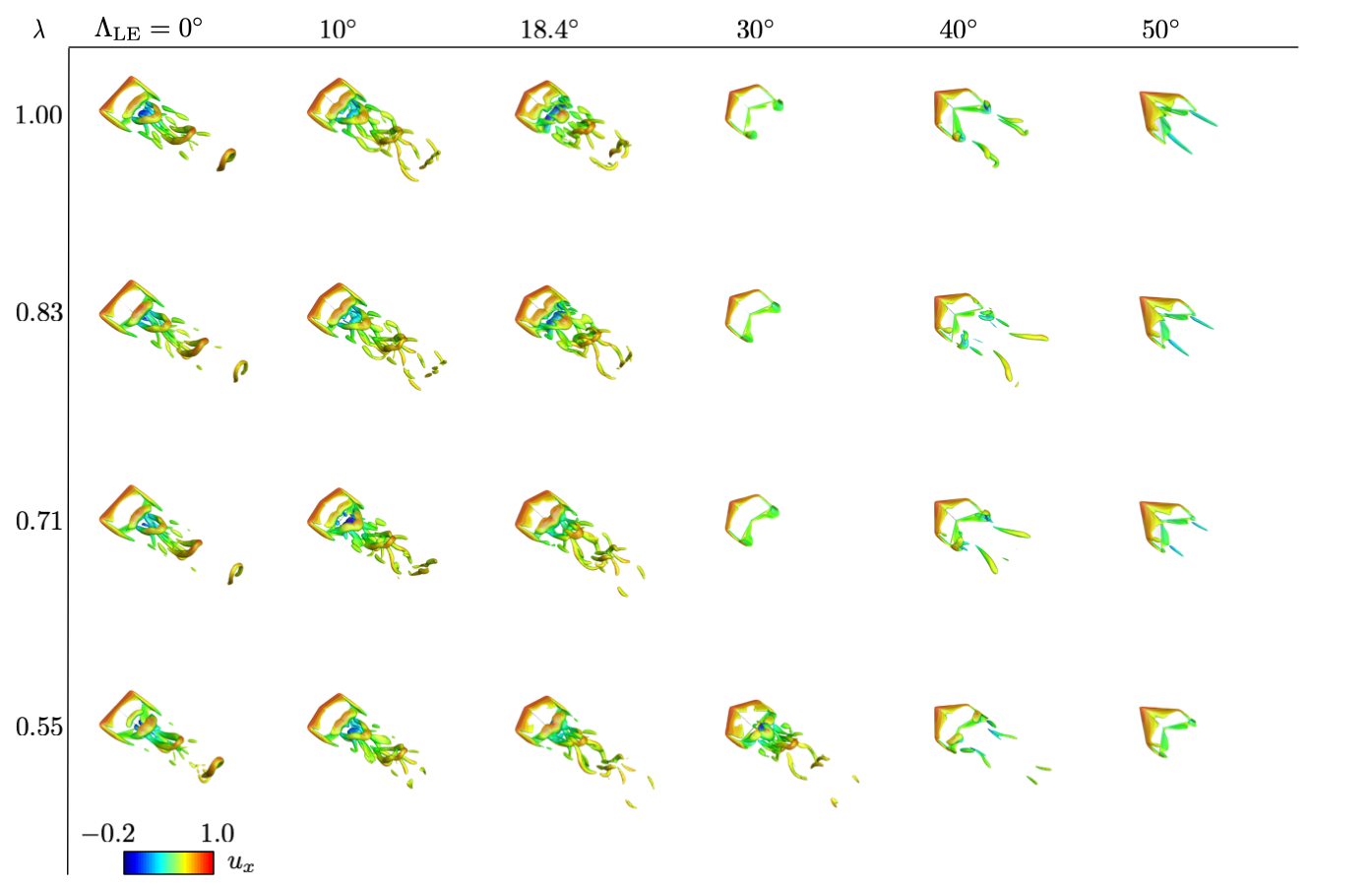}};
		\end{tikzpicture} \vspace{0mm}
		\caption{Instantaneous flow fields around wings of $sAR = 1$ at $\alpha = 22^\circ$ visualized using isosurfaces of $Q = 1$ colored by streamwise velocity $u_x$.} 
		\label{fig:perspective_ar1aoa22}
	\end{figure}

\begin{figure}
		\centering
		\begin{tikzpicture}
		\node[anchor=south west,inner sep=0] (image) at (0,0) {\includegraphics[trim=0mm 0mm 0mm 0mm, clip,width=1\textwidth]{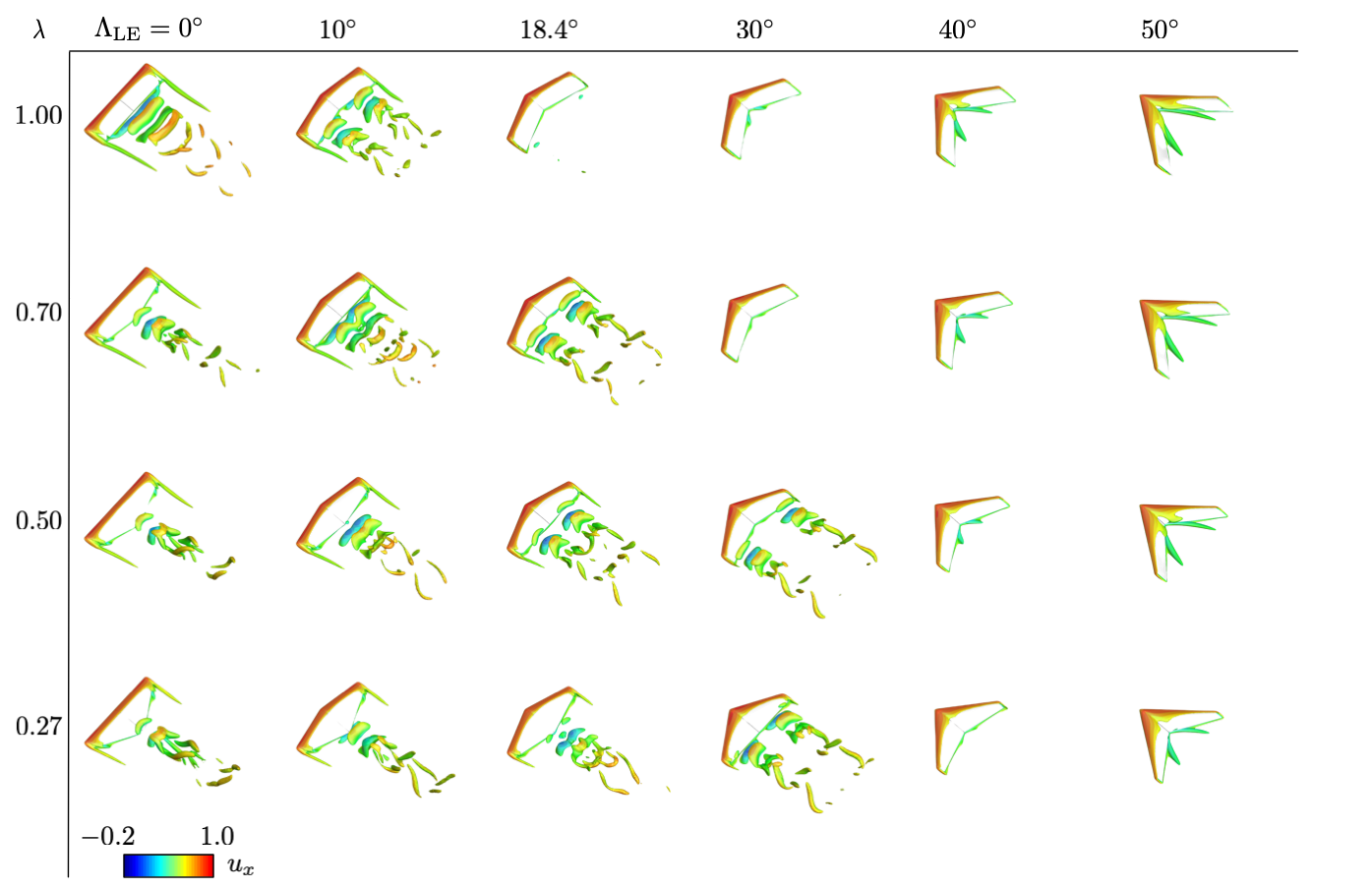}};
		\end{tikzpicture} \vspace{0mm}
		\caption{Instantaneous flow fields around wings of $sAR = 2$ at $\alpha = 14^\circ$ visualized using isosurfaces of $Q = 1$ colored by streamwise velocity $u_x$.} 
		\label{fig:perspective_ar2aoa14}
	\end{figure}

\begin{figure}
		\centering
		\begin{tikzpicture}
		\node[anchor=south west,inner sep=0] (image) at (0,0) {\includegraphics[trim=0mm 0mm 0mm 0mm, clip,width=1\textwidth]{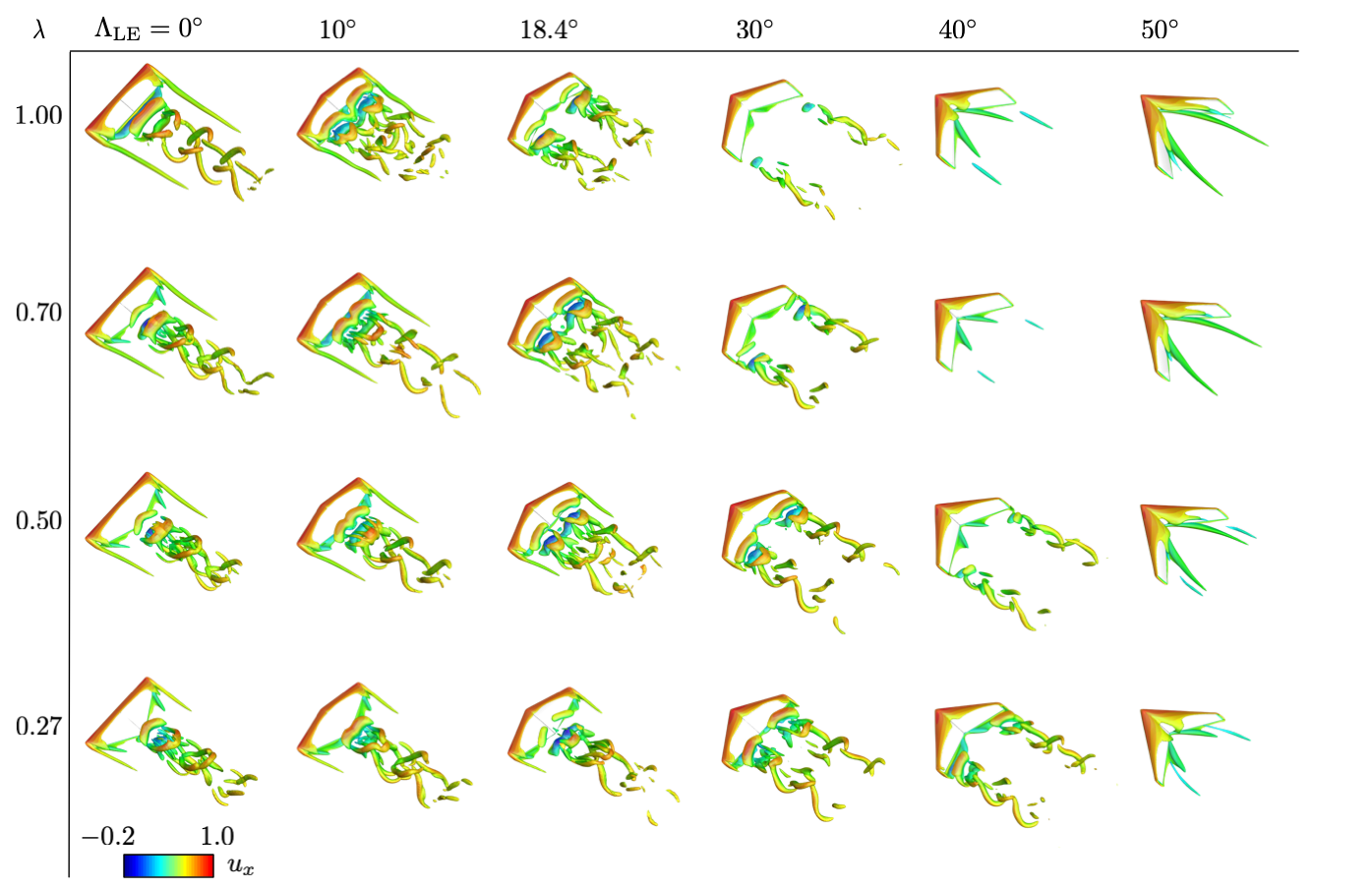}};
		\end{tikzpicture} \vspace{0mm}
		\caption{Instantaneous flow fields around wings of $sAR = 2$ at $\alpha = 18^\circ$ visualized using isosurfaces of $Q = 1$ colored by streamwise velocity $u_x$.} 
		\label{fig:perspective_ar2aoa18}
	\end{figure}

\begin{figure}
		\centering
		\begin{tikzpicture}
		\node[anchor=south west,inner sep=0] (image) at (0,0) {\includegraphics[trim=0mm 0mm 0mm 0mm, clip,width=1\textwidth]{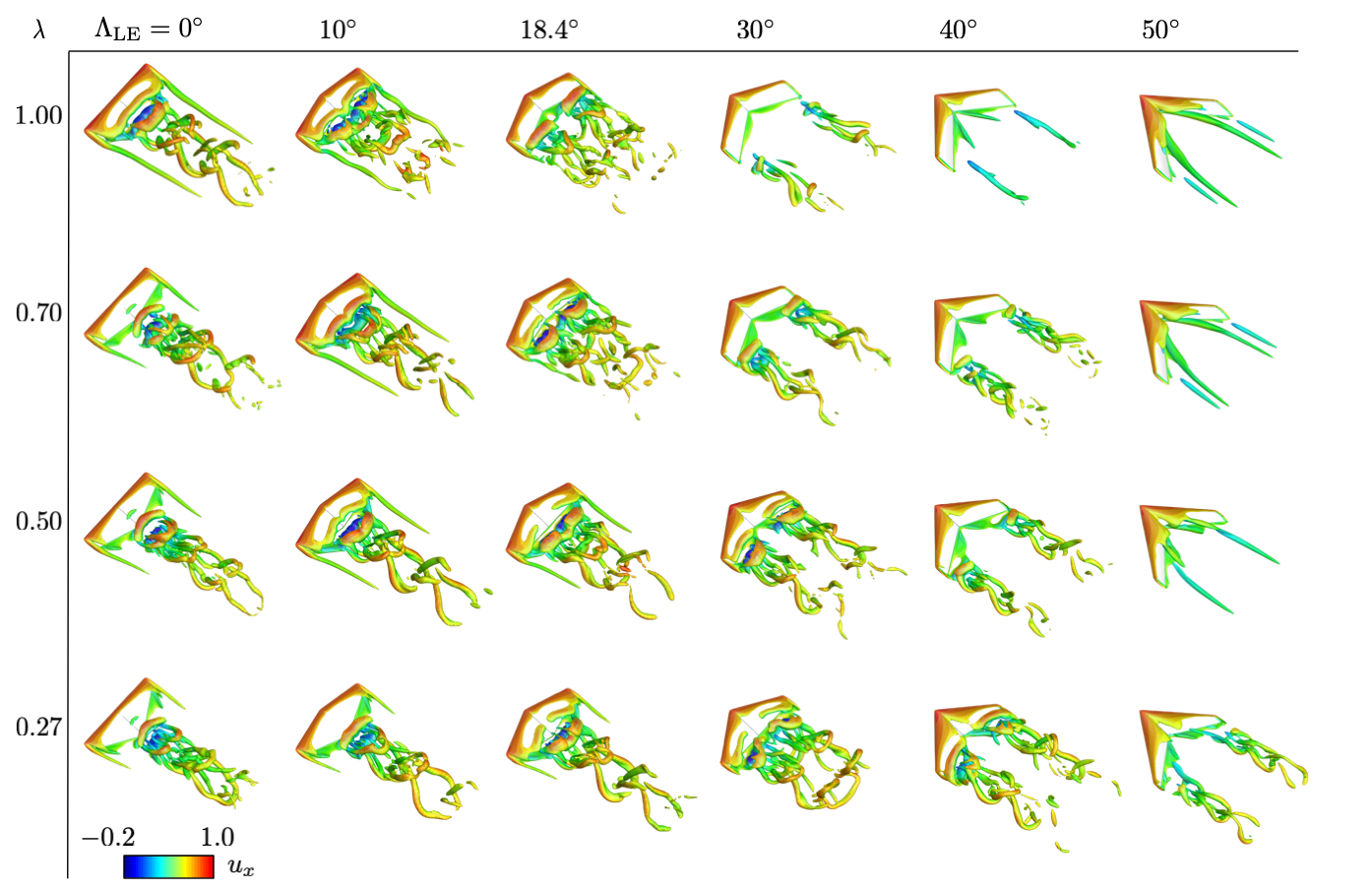}};
		\end{tikzpicture} \vspace{0mm}
		\caption{Instantaneous flow fields around wings of $sAR = 2$ at $\alpha = 22^\circ$ visualized using isosurfaces of $Q = 1$ colored by streamwise velocity $u_x$.} 
		\label{fig:perspective_ar2aoa22}
	\end{figure}

	In this appendix, we provide flow field visualizations of the wake structures around all tapered wings considered in the present study. Flows around $sAR = 1$ wings at $\alpha = 14^\circ$, $18^\circ$, and $22^\circ$ are shown in figures $\ref{fig:perspective_ar1aoa14}$, \ref{fig:perspective_ar1aoa18}, and \ref{fig:perspective_ar1aoa22}, respectively. 
	Similarly,  flows around $sAR = 2$ wings $\alpha = 14^\circ$, $18^\circ$, and $22^\circ$ are shown in figures \ref{fig:perspective_ar2aoa14}, \ref{fig:perspective_ar2aoa18}, and \ref{fig:perspective_ar2aoa22}, respectively.
	All flows are visualized using isosurfaces of $Q = 1$, colored by the streamwise velocity $u_x$.
	
	
	\section{Force element analysis}
	\label{sec:forceleements}
	Force element analysis \citep{Chang:PRSA92} is a method used to identify vortical structures emerging near the wake that play a significant role in exerting aerodynamic loads over the wing. We note that force element theory is similar in spirit to other force element methods derived through a variational approach \citep{Quartapelle:AIAAJ83}, as well as to vortex force maps \citep{Li:JFM20}, and  force partition approaches \citep	{Menon:JCP21}. The present force element analysis was previously employed to analyze incompressible flows over wings in multiple configurations \citep{Lee:JFM12,Zhang:JFM20,Zhang:AIAAJ22,Zhang:PRF22}. 
	
	Initially, we define an auxiliary potential with a specific boundary condition of $-\mathbi{n} \cdot \nabla \phi_i = \mathbi{n} \cdot \mathbi{e}_i$ set on the wing surface, where $\phi$ is the auxiliary potential, $\mathbi{n}$ is the unit wall normal vector, and $\mathbi{e}_i$ is the unit vector in the $i$th direction. The inner product of the Navier--Stokes equations with $\nabla \phi$ and performing an integral over the fluid domain, the forces exerted in the $i$-th direction may be expressed as
	\begin{equation}
	F_i = \int_V \bs{\omega} \times \mathbi{u} \cdot \nabla \phi_i \text{d}V + \frac{1}{Re} \int_S \bs{\omega} \times \mathbi{n} \cdot (\nabla \phi_i + \mathbi{e_i}) \text{d}S,
	\label{eq:6_forceelements}
	\end{equation}
	where the first integral represents the volume force elements and the second integral term is comprised of the surface force elements. For $Re_c = 600$ flows, the volume elements  are responsible for the major contribution to the total force over the wing. We note that the auxiliary potential velocity field $\phi_i$ decays rapidly far from the surface. For this reason, the structures that have a higher contribution to lift are the ones that emerge near the surface. 
	
	To visualize the vortical structures associated with lift generation, one can take the Hadamard product of the  $\nabla \phi_i$ and the Lamb vector ($\bs{\omega} \times \mathbi{u}$), as shown in figure \ref{fig:ForceElements}. The resulting $(\mathbf{u} \times \bs{\omega}) \cdot \nabla \phi_y$ variable is often called as lift element. Lastly, we recall that the force element theory used herein considers incompressible Navier--Stokes equations. Nevertheless, important insights on the flow structures may still be obtained for weakly compressible flows, such as the ones considered herein and in previous studies as well \citep{Ribeiro:JFM22}.
	
	\section*{Acknowledgments}
	\label{sec:acknowledgments}
	We acknowledge the support from the US Air Force Office of Scientific Research (program manager: Dr. G. Abate, grant: FA9550-21-1-0174). The first author thanks T.  R. Ricciardi and K. Zhang for the enlightening discussions. Computational resources were provided by the High Performance Computing Modernization Program at the US Department of Defense and the Texas Advanced Computing Center. 
	
	\section*{Declaration of interest}
	\label{sec:doi}
	The authors report no conflict of interest.
	
	\bibliography{taira_refs}
	\bibliographystyle{jfm}
	
\end{document}